\documentclass[a4paper,12pt]{article}
\usepackage{amssymb,graphics,a4wide}
\usepackage[intlimits,fleqn]{amsmath}
\numberwithin{equation}{section}
\newcommand{\MS}{\ensuremath{\overline{\mathrm{MS}}}}
\DeclareMathOperator{\Tr}{Tr}
\DeclareMathOperator{\T}{T}

\begin{document}

\begin{flushright}
\Large TTP03-27\\
SFB/CPP-03-49
\end{flushright}
\vspace{10mm}
\begin{center}
\Huge Renormalons:\\
technical introduction\\[10mm]
\Large Andrey Grozin\\
Institut f\"ur Theoretische Teilchenphysik,
Universit\"at Karlsruhe\\
grozin@particle.uni-karlsruhe.de
\end{center}

\begin{abstract}
The large-$\beta_0$ limit of QCD is discussed,
with the emphasize on simple technical methods of calculating
various quantities at the order $1/\beta_0$.
Many examples, mainly from heavy quark physics,
are considered.
Some QCD results based on renormalization group
(and not restricted to the large-$\beta_0$ limit)
are also discussed.
\end{abstract}

\section{Large-$\beta_0$ Limit}
\label{Sec:beta}

It is well known that perturbative series do not converge.
They are asymptotic series,
i.e., the difference between the exact result
and its approximation up to the order $\alpha_s^L$,
divided by $\alpha_s^L$, tends to 0 in the limit $\alpha_s\to0$.
Large-order behaviour of various perturbative series
attracted considerable attention during recent years.
Most of the results obtained so far are model-dependent:
they are derived in the large-$\beta_0$ limit, i.e., at $n_f\to-\infty$.
There are some hints that the situation in the real QCD
may be not too different from this limit, but this cannot be proved.
However, a few results are rigorous consequences of QCD;
they are based on the renormalization group.
Many more applications are discussed in the excellent review~\cite{Be:99},
where additional references can be found.

Let's consider a perturbative quantity $A$
such that the tree diagram for it contains no gluon propagators.
We can always normalize the tree value of $A$ to be 1.
Then the perturbative series for the bare quantity $A_0$ has the form
\begin{equation}
A_0 = 1 + \sum_{L=1}^\infty \sum_{n=0}^{L-1} a'_{Ln} n_f^n
\left(\frac{g_0^2}{(4\pi)^{d/2}}\right)^L\,,
\label{beta:form0}
\end{equation}
where $L$ is the number of loops.
This series can be rewritten in terms of
$\beta_0=\frac{11}{3}C_A-\frac{4}{3}T_F n_f$ instead of $n_f$:
\begin{equation}
A_0 = 1 + \sum_{L=1}^\infty \sum_{n=0}^{L-1} a_{Ln} \beta_0^n
\left(\frac{g_0^2}{(4\pi)^{d/2}}\right)^L\,.
\label{beta:form1}
\end{equation}
Now we are going to consider $\beta_0$ as a large parameter
such that $\beta_0\alpha_s\sim1$,
and consider only a few terms in the expansion in $1/\beta_0\sim\alpha_s$:
\begin{equation}
A_0 = 1 + \frac{1}{\beta_0} f\left(\frac{\beta_0 g_0^2}{(4\pi)^{d/2}}\right)
+ \mathcal{O}\left(\frac{1}{\beta_0^2}\right)\,.
\label{beta:form2}
\end{equation}
This regime is called large-$\beta_0$ limit,
and can only hold in QCD with $n_f\to-\infty$.
Note that it has nothing in common with the large-$N_c$ limit,
because we cannot control powers of $N_c$ in the coefficients $a_{Ln}$.

There is some empirical evidence~\cite{BG:95}
that the two-loop coefficients $a_{21}\beta_0+a_{20}$
for many quantities are well approximated by $a_{21}\beta_0$.
It is easy to find $a'_{21}$ from the diagram with the quark-loop insertion
into the gluon propagator in the one-loop correction.
Then we can estimate the full two-loop coefficient as
$a_{21}\beta_0=a'_{21}\left(n_f-\frac{11}{4}C_A/T_F\right)$.
This is called naive nonabelianization~\cite{BG:95}.
Of course, there is no guarantee that this will hold at higher orders.
We can only hope that higher perturbative corrections are mainly
due to the running of $\alpha_s$;
in this respect, gluonic contributions behave as $-33/2$ flavours,
and QCD with $n_f=3$ or $4$ flavours is not too different from
QCD with $-\infty$ flavours.

It is easy to find the coefficients $a_{L,L-1}$
of the highest degree $\beta_0^{L-1}$ at $L$ loops.
They are determined by the coefficients $a'_{L,L-1}$ of $n_f^{L-1}$,
i.e., by inserting $L-1$ quark loops into the gluon propagator
in the one-loop correction.
We shall assume for now that there is only one gluon propagator
and no three-gluon vertices at one loop.
The bare gluon propagator with $L-1$ quark loops inserted is
\begin{equation}
\begin{split}
D_{\mu\nu}^{(L-1)}(p) ={}& - \frac{1}{(-p^2)^{1+(L-1)\varepsilon}}
\left(g_{\mu\nu}+\frac{p_\mu p_\nu}{-p^2}\right)
\left(-\frac{4}{3} T_F n_f \frac{g_0^2}{(4\pi)^{d/2}}
\frac{D(\varepsilon)}{\varepsilon} e^{-\gamma\varepsilon}\right)^{L-1}\,,\\
D(\varepsilon) ={}& 6 e^{\gamma\varepsilon} \Gamma(1+\varepsilon)
B(2-\varepsilon,2-\varepsilon) = 1 + \frac{5}{3} \varepsilon + \cdots
\end{split}
\label{beta:D}
\end{equation}
It looks like the free propagator
\begin{equation*}
D_{\mu\nu}(p) = - \frac{1}{-p^2}
\left( g_{\mu\nu} + (1-a_0) \frac{p_\mu p_\nu}{-p^2} \right)
\end{equation*}
in the Landau gauge $a_0=0$,
with a shifted power of $-p^2$ (and an extra constant factor).
Let the one-loop contribution to $A_0$ be
$\left(a_1+a'_1 a_0\right) g_0^2/(4\pi)^{d/2}$.
If we calculate the one-loop contribution in the Landau gauge $a_1$
with the denominator of the gluon propagator equal to $(-p^2)^n$
instead of just $-p^2$ and call it $a_1(n)$,
then for all $L>1$
\begin{equation}
a_{L,L-1} =
\left(\frac{D(\varepsilon)}{\varepsilon}e^{-\gamma\varepsilon}\right)^{L-1}
a_1\bigl(1+(L-1)\varepsilon\bigr)\,.
\label{beta:aL}
\end{equation}
Only the one-loop contribution $a_{10}$ contains the additional
gauge-dependent term $a'_1 a_0$.

The large-$\beta_0$ limit, as formulated above,
does not correspond to summation of any subset of diagrams.
If we include not only quark loops, but also gluon and ghost ones,
then $-\frac{4}{3}T_F n_f$ in~(\ref{beta:D}) is replaced by
\begin{equation}
\beta_0 - \frac{C_A}{3}
\left\{8\varepsilon + \frac{3-2\varepsilon}{2(1-\varepsilon)} (a_0+3)
\left[1 - \frac{\varepsilon}{2} (a_0+3)\right] \right\}\,.
\label{beta:AntiYennie}
\end{equation}
Summing these diagrams yields a gauge-dependent result.
This gauge dependence is compensated (for a gauge-invariant $A_0$)
by other diagrams, which have more complicated topologies than
a simple chain, and are impossible to sum.
In the gauge $a_0=-3$, one-loop running of $\alpha_s$ is produced by
one-loop insertions in the gluon propagator only,
without vertex contributions.
Summation of chains of one-loop insertions into the gluon propagator
in this gauge is equivalent to the large-$\beta_0$ limit.

In the large-$\beta_0$ limit, $\beta_1\sim\beta_0$, $\beta_2\sim\beta_0^2$, etc.
Therefore, $\beta$-function is equal to
\begin{equation}
\beta = \frac{\beta_0\alpha_s}{4\pi}
\label{beta:beta}
\end{equation}
(this term is of order 1) plus $\mathcal{O}(1/\beta_0)$ corrections.
At the leading order in $1/\beta_0$, the renormalization-group equation
\begin{equation}
\frac{d\log Z_\alpha}{d\beta} = - \frac{\beta}{\varepsilon+\beta}
\label{beta:RG}
\end{equation}
can be explicitly integrated:
\begin{equation}
Z_\alpha = \frac{1}{1+\beta/\varepsilon}\,.
\label{beta:Za}
\end{equation}
To the leading order in $1/\beta_0$,
\begin{equation}
\alpha_s(\mu) = \frac{2\pi}{\beta_0\log\frac{\mu}{\Lambda_{\MS}}}\,.
\label{beta:as}
\end{equation}

The perturbative series~(\ref{beta:form1}) can be rewritten (in the Landau gauge)
via the renormalized quantities:
\begin{equation}
A_0 = 1 + \frac{1}{\beta_0} \sum_{L=1}^\infty \frac{F(\varepsilon,L\varepsilon)}{L}
\left(\frac{\beta}{\varepsilon+\beta}\right)^L
+ \mathcal{O}\left(\frac{1}{\beta_0^2}\right)\,,
\label{beta:formF}
\end{equation}
where
\begin{equation}
F(\varepsilon,u) = u e^{\gamma\varepsilon} a_1(1+u-\varepsilon)
\mu^{2u} D(\varepsilon)^{u/\varepsilon-1}\,.
\label{beta:F}
\end{equation}
If $a\ne0$, the term $a'_1 a \beta/\beta_0 + \mathcal{O}(1/\beta_0^2)$
should be added
(the difference between $a_0$ and $a$ is $\mathcal{O}(1/\beta_0)$).

We can expand~(\ref{beta:formF}) in the renormalized $\alpha_s$,
or in $\beta$~(\ref{beta:beta}), using
\begin{equation*}
\left(\frac{\beta}{\varepsilon+\beta}\right)^L =
\left(\frac{\beta}{\varepsilon}\right)^L \left[1 - L \frac{\beta}{\varepsilon}
+ \frac{(L)_2}{2} \left(\frac{\beta}{\varepsilon}\right)^2
- \frac{(L)_3}{3!} \left(\frac{\beta}{\varepsilon}\right)^3 + \cdots\right]
\end{equation*}
(here $(x)_n=x(x+1)\cdots(x+n-1)=\Gamma(x+n)/\Gamma(x)$ is the Pochhammer symbol).
In the applications we shall consider,
$F(\varepsilon,u)$ is regular at the origin:
\begin{equation}
F(\varepsilon,u)=\sum_{n=0}^\infty\sum_{m=0}^\infty F_{nm}\varepsilon^n u^m\,,
\label{beta:Fexp}
\end{equation}
though I know no general proof of this fact.
Substituting these expansions to~(\ref{beta:formF}),
we obtain a quadruple sum expressing $A_0$ via the renormalized quantities.

The bare quantity $A_0=Z A$, where both $Z$ and $A$ have the form
$1+\mathcal{O}(1/\beta_0)$.
Therefore, we can find $Z-1$ with the $1/\beta_0$ accuracy
just by retaining all terms with negative powers of $\varepsilon$
in this quadruple sum.
The renormalized $A-1$, with the $1/\beta_0$ accuracy,
is given by terms with $\varepsilon^0$.
It is enough to find $Z_1$, the coefficient of $1/\varepsilon$ in $Z$,
in order to have the anomalous dimension
\begin{equation*}
\gamma = -2 \frac{d Z_1}{d\log\beta}\,.
\end{equation*}
Collecting terms with $\varepsilon^{-1}$ in the quadruple sum for $A_0$,
we obtain for $\beta_0 Z_1$
\begin{gather*}
\beta F_{00} - \beta^2 (F_{10}+F_{01}) + \beta^3 (F_{20}+F_{11}+F_{02})
- \beta^4 (F_{30}+F_{21}+F_{12}+F_{03}) + \cdots\\
{} + \tfrac{1}{2} \beta^2 (F_{10}+2F_{01}) - \beta^3 (F_{20}+2F_{11}+4F_{02})
+ \tfrac{3}{2} \beta^4 (F_{30}+2F_{21}+4F_{12}+8F_{03}) + \cdots\\
{} +\tfrac{1}{3} \beta^3 (F_{20}+3F_{11}+9F_{02})
- \beta^4 (F_{30}+3F_{21}+9F_{12}+27F_{03}) + \cdots\\
{} + \tfrac{1}{4} \beta^4 (F_{30}+4F_{21}+16F_{12}+64F_{03}) + \cdots\\
{} + \cdots\\
= \beta F_{00} - \frac{\beta^2}{2} F_{10}
+ \frac{\beta^3}{3} F_{20} - \frac{\beta^4}{4} F_{30} + \cdots
\end{gather*}
Therefore, the anomalous dimension is~\cite{PP:84}
\begin{equation}
\gamma = - 2 \frac{\beta}{\beta_0} F(-\beta,0)
+ \mathcal{O}\left(\frac{1}{\beta_0^2}\right)\,.
\label{beta:gamma}
\end{equation}

Collecting terms with $\varepsilon^0$ in the quadruple sum for $A_0$,
we obtain for $\beta_0 (A-1)$
\begin{gather*}
\beta (F_{10}+F_{01})
- \beta^2 (F_{20}+F_{11}+F_{02})
+ \beta^3 (F_{30}+F_{21}+F_{12}+F_{03})\\
{}\qquad\qquad{} - \beta^4 (F_{40}+F_{31}+F_{22}+F_{13}+F_{04}) + \cdots\\
{} + \tfrac{1}{2} \beta^2 (F_{20}+2F_{11}+4F_{02})
- \beta^3 (F_{30}+2F_{21}+4F_{12}+8F_{03})\\
{}\qquad\qquad{}
+ \tfrac{3}{2} \beta^4 (F_{40}+2F_{31}+4F_{22}+8F_{13}+16F_{04}) + \cdots\displaybreak\\
{} + \tfrac{1}{3} \beta^3 (F_{30}+3F_{21}+9F_{12}+27F_{03})
- \beta^4 (F_{40}+3F_{31}+9F_{22}+27F_{13}+81F_{04}) + \cdots\\
{} + \tfrac{1}{4} \beta^4 (F_{40}+4F_{31}+16F_{22}+64F_{13}+256F_{04}) + \cdots\\
{} + \cdots\\
{} = \beta F_{10} - \frac{\beta^2}{2} F_{20} + \frac{\beta^3}{3} F_{30}
- \frac{\beta^4}{4} F_{40} + \cdots\\
\hphantom{{}={}} + \beta F_{01} + \beta^2 F_{02} + 2 \beta^3 F_{03}
+ 6 \beta^4 F_{04} + \cdots
\end{gather*}
Therefore, the renormalized quantity is~\cite{Br:93}
\begin{equation}
A(\mu) = 1 + \frac{1}{\beta_0} \int_{-\beta}^0 d\varepsilon
\frac{F(\varepsilon,0)-F(0,0)}{\varepsilon}
+ \frac{1}{\beta_0} \int_0^\infty d u\,e^{-u/\beta}
\frac{F(0,u)-F(0,0)}{u}
+ \mathcal{O}\left(\frac{1}{\beta_0^2}\right)\,,
\label{beta:A}
\end{equation}
where $\beta=\beta_0\alpha_s(\mu)/(4\pi)$.

The renormalization group equation
\begin{equation*}
\frac{d\log A(\mu)}{d\log\alpha_s} = \frac{\gamma(\alpha_s)}{2\beta(\alpha_s)}
\end{equation*}
can be conveniently solved as
\begin{equation}
A(\mu) = \hat{A}
\left(\frac{\alpha_s(\mu)}{\alpha_s(\mu_0)}\right)^{\frac{\gamma_0}{2\beta_0}}
K_\gamma(\alpha_s(\mu))\,,
\label{beta:RGsol}
\end{equation}
where the function
\begin{equation}
K_\gamma(\alpha_s) = \exp \int_0^{\alpha_s}
\left( \frac{\gamma(\alpha_s)}{2\beta(\alpha_s)} - \frac{\gamma_0}{2\beta_0} \right)
\frac{d\alpha_s}{\alpha_s} =
1 + \frac{\gamma_0}{2\beta_0}
\left( \frac{\gamma_1}{\gamma_0} - \frac{\beta_1}{\beta_0} \right)
\frac{\alpha_s}{4\pi} + \cdots
\label{CMag:K}
\end{equation}
satisfying
\begin{equation*}
K_0(\alpha_s) = 1\,,\quad
K_{-\gamma}(\alpha_s) = K^{-1}_\gamma(\alpha_s)\,,\quad
K_{\gamma_1+\gamma_2}(\alpha_s) = K_{\gamma_1}(\alpha_s) K_{\gamma_2}(\alpha_s)
\end{equation*}
has been introduced, and
\begin{equation*}
\hat{A} = A(\mu_0) K_{-\gamma}(\alpha_s(\mu_0))\,.
\end{equation*}
At the first order in $1/\beta_0$, we obtain from~(\ref{beta:gamma})
\begin{equation*}
K_\gamma(\alpha_s) = 1 + \frac{1}{\beta_0} \int_{-\beta(\alpha_s)}^0 d\varepsilon
\frac{F(\varepsilon,0)-F(0,0)}{\varepsilon}\,.
\end{equation*}
Therefore,
\begin{equation*}
\hat{A} = 1 + \frac{1}{\beta_0} \int_0^\infty d u\,e^{-u/\beta(\alpha_s(\mu_0))}
\left.\frac{F(0,u)-F(0,0)}{u}\right|_{\mu_0}
+ \mathcal{O}\left(\frac{1}{\beta_0^2}\right)\,.
\end{equation*}

Let's suppose that $m\gg\Lambda_{\MS}$ is the characteristic
hard scale in the quantity $A$.
Then $F(\varepsilon,u)$ contains the factor $(\mu/m)^{2u}$.
When taking the limit $\varepsilon\to0$,
the factor $D(\varepsilon)^{u/\varepsilon-1}$ in~(\ref{beta:F})
becomes $\exp\left(\frac{5}{3}u\right)$.
Therefore,
\begin{equation}
F(0,u) = \left(\frac{e^{5/6}\mu}{m}\right)^{2u} F(u)\,,\quad
F(u) = u a_1(1+u) m^{2u}\,.
\label{beta:Fu}
\end{equation}
It is most convenient to use
\begin{equation}
\mu_0=e^{-5/6}m
\label{beta:mu0}
\end{equation}
in the definition of $\hat{A}$.
In the rest of this Chapter, $\beta$ will mean $\beta_0\alpha_s(\mu_0)/(4\pi)$.
This renormalization-group invariant is
\begin{equation}
\hat{A} = 1 + \frac{1}{\beta_0} \int_0^\infty d u\,e^{-u/\beta} S(u)
+ \mathcal{O}\left(\frac{1}{\beta_0^2}\right)\,,
\label{beta:Ahat}
\end{equation}
where
\begin{equation}
S(u) = \frac{F(u)-F(0)}{u}\,.
\label{beta:Su}
\end{equation}
Here,
\begin{equation}
e^{-u/\beta} = \left(\frac{e^{5/6} \Lambda_{\MS}}{m}\right)^{2u}\,.
\label{beta:mcor}
\end{equation}

If we substitute the expansion
\begin{equation*}
S(u) = \sum_{L=1}^\infty s_L u^{L-1}
\end{equation*}
into the Laplace integral~(\ref{beta:Ahat}),
we obtain the renormalized perturbative series
\begin{gather}
\hat{A} = 1 + \frac{1}{\beta_0} \sum_{L=1}^\infty c_L \beta^L
+ \mathcal{O}\left(\frac{1}{\beta_0^2}\right)\,,
\label{beta:rseries}\\
c_L = (L-1)!\,s_L = \left.\left(\frac{d}{d u}\right)^{L-1} S(u)\right|_{u=0}\,.
\label{beta:cL}
\end{gather}
Therefore, $S(u)$ can be obtained from $\hat{A}$~(\ref{beta:rseries}) by
\begin{equation}
S(u) = \sum_{L=1}^\infty \frac{c_L u^{L-1}}{(L-1)!}\,,
\label{beta:Borel}
\end{equation}
which is called Borel transform.

We see that the function $F(\varepsilon,u)$~(\ref{beta:F})
contains all the necessary information about the quantity $A$
at the $1/\beta_0$ order.
The anomalous dimension~(\ref{beta:gamma}) is determined by $F(\varepsilon,0)$,
and the renormalization-group invariant $\hat{A}$~(\ref{beta:Ahat})
(which gives $A(\mu)$~(\ref{beta:RGsol})) -- by $F(0,u)$.
These formulae are written in the Landau gauge $a=0$;
if $a\ne0$, additional one-loop terms from the longitudinal part
of the gluon propagator should be added.

\section{Renormalons}
\label{Sec:Ren}

The Laplace integral~(\ref{beta:Ahat}) is not well-defined
if the Borel image $S(u)$ has singularities on the integration path --
the positive half-axis $u>0$.
At the first order in $1/\beta_0$, $S(u)$ typically has simple poles.
If
\begin{equation}
S(u) = \frac{r}{u_0-u} + \cdots
\label{Ren:Pole}
\end{equation}
where dots mean terms regular at $u=u_0$,
and $u_0>0$, then the integral~(\ref{beta:Ahat}) is not well-defined near $u_0$.
One way to make sense of this integral is to use its principal value:
to make a hole $[u_0-\delta,u_0+\delta]$ and take the limit $\delta\to0$.
However, if we make, e.g., a hole $[u_0-\delta,u_0+2\delta]$ instead,
we'll get a result which differs from the principal value by the residue
of the integrand times $\log2$.
Therefore, the sum of the perturbative series~(\ref{beta:Ahat})
contains an intrinsic ambiguity of the order of this residue.
It is equal to
\begin{equation}
\Delta \hat{A} = \frac{r e^{-u_0/\beta}}{\beta_0}
= \frac{r}{\beta_0} \left(\frac{e^{5/6} \Lambda_{\MS}}{m}\right)^{2u_0}\,.
\label{Ren:Ambig}
\end{equation}
These renormalon ambiguities are commensurate with $1/m$ power corrections --
contributions of matrix elements of higher-dimensional operators
to the quantity $A$.
The full result for the physical quantity $A$ must be unambiguous.
Therefore, if one changes prescription for handling the integral
across the renormalon singularity at $u=u_0$,
one has to change the values of the dimension-$2u_0$ matrix elements
accordingly.
This shows that renormalons can only happen at integer and half-integer
values of $u$, corresponding to dimensionalities of allowed power corrections.
The largest ambiguity is associated with the renormalon closest to the origin.

The renormalon pole~(\ref{Ren:Pole}) yields the contribution
to the coefficients $c_L$ of the renormalized
perturbative series~(\ref{beta:rseries}) equal to
\begin{equation}
c_L = r \frac{(L-1)!}{u_0^L}
\label{Ren:asympt}
\end{equation}
(see~(\ref{beta:cL})).
The series~(\ref{Ren:Pole}) is, clearly, divergent.
Using the Stirling formula for the factorial, we can see that
the terms of this series behave as
\begin{equation*}
c_L \beta^L \sim r \left(\frac{\beta L}{e u_0}\right)^L
\end{equation*}
at large $L$.
The best one can do with such a series is to sum it until its minimum term,
and to assign it an ambiguity of the order of this minimum term.
The minimum happens at $L\approx u_0/\beta$ loops,
and the magnitude of the minimum term is given by~(\ref{Ren:Ambig}).
This is another way to look at this renormalon ambiguity.
The fastest-growing contribution to $c_L$ comes from the renormalon
most close to the origin.

Note that renormalons at $u_0<0$ give sign-alternating factorially-growing
coefficients~(\ref{Ren:asympt}).
For such series, the integral~(\ref{beta:Ahat}) provides an unambiguous
definition of summation called Borel sum.

Renormalon singularities can result from either UV or IR divergences
of the one-loop integral.
Suppose that it behaves at $k\to\infty$
as $\int d^4 k/(-k^2)^{n_{\text{UV}}}$,
so that the degree of its UV divergence (at $d=4$)
is $\nu_{\text{UV}}=4-2n_{\text{UV}}$.
When we insert the renormalon chain, the power changes:
$n_{\text{UV}}\to n_{\text{UV}}+(L-1)\varepsilon=n_{\text{UV}}+u$
if $\varepsilon=0$ (which is the case when calculating $S(u)$).
This integral can have an UV divergence only
at $u\le2-n_{\text{UV}}=\nu_{\text{UV}}/2$.
Therefore, UV renormalons can be situated at $\nu_{\text{UV}}/2$
and to the left.
Only quantities $A$ with power-like UV divergences at one loop
have UV renormalons at positive $u$.
The divergence at $u=0$ is the usual UV divergence of the one-loop integral,
which is eliminated by renormalization;
renormalized quantities have no UV renormalon at $u=0$.

Similarly, if the one-loop integral behaves
as $\int d^4 k/(-k^2)^{n_{\text{IR}}}$ at $k\to0$
(where $k$ is the virtual gluon momentum),
so that the degree of its IR divergence is $\nu_{\text{IR}}=2n_{\text{IR}}-4$,
$S(u)$ can have an IR divergence only at $u\ge2-n=-\nu_{\text{IR}}/2$,
and IR renormalons can be situated at $-\nu_{\text{IR}}/2$
and integer and half-integer points to the right from it.
Quantities described by off-shell diagrams have $n_{\text{IR}}=1$,
and their IR renormalons are at $u=1$ and to the right.

We can get a better understanding of the physical meaning of renormalons
if we rewrite~(\ref{beta:Ahat}) in the form~\cite{Ne:95}
\begin{equation}
\hat{A} = 1 + \int_0^\infty \frac{d\tau}{\tau} w(\tau)
\frac{\alpha_s(\sqrt{\tau}\mu_0)}{4\pi}
+ \mathcal{O}\left(\frac{1}{\beta_0^2}\right)\,.
\label{Ren:Neubert}
\end{equation}
This looks like the one-loop correction,
but with the running $\alpha_s$ under the integral sign.
The function $w(\tau)$ has the meaning of the distribution function
in gluon virtualities in the one-loop diagram;
it is normalized to the coefficient of $\alpha_s/(4\pi)$
in the one-loop correction.
Inside the $1/\beta_0$ term in~(\ref{Ren:Neubert}),
we may use the leading-order formula for running of $\alpha_s$:
\begin{equation*}
\alpha_s(\sqrt{\tau}\mu_0) = \frac{\alpha_s(\mu_0)}{1+\beta\log\tau}
= \alpha_s(\mu_0) \sum_{n=0}^\infty (-\beta\log\tau)^n\,.
\end{equation*}
Substituting this expansion into~(\ref{Ren:Neubert}),
we see, that this representation holds if $w(\tau)$
is related to the perturbative series coefficients $c_L$ by
\begin{equation}
c_L = \int_0^\infty \frac{d\tau}{\tau} w(\tau) (-\log\tau)^{L-1}\,.
\label{Ren:cL}
\end{equation}
Therefore, $S(u)$~(\ref{beta:Borel}) becomes
\begin{equation}
S(u) = \int_0^\infty \frac{d\tau}{\tau} w(\tau) \tau^{-u}\,.
\label{Ren:Su}
\end{equation}
In other words, $S(u)$ is the Mellin transform of $w(\tau)$.
Therefore, the distribution function $w(\tau)$ is given by the inverse
Mellin transform:
\begin{equation}
w(\tau) = \frac{1}{2\pi i} \int_{u_0-i\infty}^{u_0+i\infty}
d u\,S(u) \tau^u\,,
\label{Ren:w}
\end{equation}
where $u_0$ should lie in the gap between IR and UV renormalons.

At $\tau<1$ we can close the integration contour to the right.
If $S(u)$ has IR renormalons $r_i/(u_i-u)$, then
\begin{equation}
w(\tau) = \sum_{\text{IR}} r_i \tau^{u_i}\,.
\label{Ren:w0}
\end{equation}
The leading term at small $\tau$ is given by the leftmost IR renormalon.
If our quantity $A$ is IR-finite at one loop, all $u_i>0$,
and $w(\tau)\to0$ at $\tau\to0$.
Similarly, at $\tau>1$ we can close the contour to the left.
If the UV renormalons are $r_i/(u-u_i)$, then
\begin{equation}
w(\tau) = \sum_{\text{UV}} r_i \tau^{u_i}\,.
\label{Ren:winf}
\end{equation}
The leading term at large $\tau$ is given by the rightmost UV renormalon.
If $A$ is UV-finite at one loop, all $u_i<0$,
and $w(\tau)\to0$ at $\tau\to\infty$.

All virtualities (including small ones) contribute to~(\ref{Ren:Neubert}).
Behaviour of the distribution function $w(\tau)$
in the small-virtuality region $\tau\to0$ is determined by
the IR renormalon most close to the origin.
The integral~(\ref{Ren:Neubert}) is ill-defined,
just like the original integral~(\ref{beta:Ahat}).
The one-loop $\alpha_s$~(\ref{beta:as}) becomes infinite at
$\tau=(e^{5/6}\Lambda_{\MS}/m)^2$ (Landau pole),
and we integrate across this pole.
This happens at small $\tau$; substituting the asymptotics~(\ref{Ren:w0})
of the distribution function at small virtualities,
we see that the residue at this pole,
given by the IR renormalon nearest to the origin,
is again equal to~(\ref{Ren:Ambig}).

\section{Light Quarks}
\label{Sec:RL}

First, we shall discuss the massless quark propagator
at the order $1/\beta_0$.
The one-loop expression for the quark self-energy $\Sigma(p)$
in the Landau gauge with the gluon denominator raised to the power $n=1+(L-1)\varepsilon$ is
\begin{equation*}
a_1(n) = i \frac{C_F}{-p^2} \int \frac{d^d k}{\pi^{d/2}}
\frac{\frac{1}{4}\Tr\rlap/p\gamma^\mu(\rlap/k+\rlap/p)\gamma^\nu}
{\left[-(k+p)^2\right] (-k^2)^n}
\left(g_{\mu\nu} + \frac{k_\mu k_\nu}{-k^2}\right)\,.
\end{equation*}
Using the one-loop integrals
\begin{equation}
\begin{split}
&\int \frac{d^d k}{[-(k+p)^2]^{n_1} [-k^2]^{n_2}} =
i \pi^{d/2} (-p^2)^{d/2-n_1-n_2} G(n_1,n_2)\,,\\
&G(n_1,n_2) = \frac{\Gamma(-d/2+n_1+n_2) \Gamma(d/2-n_1) \Gamma(d/2-n_2)}%
{\Gamma(n_1) \Gamma(n_2) \Gamma(d-n_1-n_2)}\,,
\end{split}
\label{RL:G1}
\end{equation}
we can easily find the function $F(\varepsilon,u)$ (\ref{beta:F}).
Such functions for all off-shell massless quantities have
the same $\Gamma$-function structure resulting from~(\ref{RL:G1})
with $n_2=1+u-\varepsilon$:
\begin{equation}
F(\varepsilon,u) = \left(\frac{\mu^2}{-p^2}\right)^u e^{\gamma\varepsilon}
\frac{\Gamma(1+u) \Gamma(1-u) \Gamma(2-\varepsilon)}
{\Gamma(2+u-\varepsilon) \Gamma(3-u-\varepsilon)}
D(\varepsilon)^{u/\varepsilon-1} N(\varepsilon,u)\,.
\label{RL:F}
\end{equation}
The first $\Gamma$-function in the numerator,
with the positive sign in front of $u$,
comes from the first $\Gamma$-function in the numerator of~(\ref{RL:G1}),
with the negative sign in front of $d$,
and its poles are UV divergences.
The second $\Gamma$-function in the numerator,
with the negative sign in front of $u$,
comes from the second $\Gamma$-function in the numerator of~(\ref{RL:G1}),
with the positive sign in front of $d$,
and its poles are IR divergences.
For $\Sigma(p)$, we obtain
\begin{equation}
N(\varepsilon,u) = - C_F (3-2\varepsilon) (u-\varepsilon)\,.
\label{RL:N}
\end{equation}
At one loop ($L=u/\varepsilon=1$),
the Landau-gauge self-energy vanishes;
at $L=2$, the $\beta_0$-term in the two-loop result is reproduced.

The massless-quark propagator $S(p)$
with the $1/\beta_0$ accuracy in the Landau gauge
is equal to $1/\rlap/p$ times~(\ref{beta:formF}),
where $F(\varepsilon,u)$ is given by~(\ref{RL:F}), (\ref{RL:N}).
Terms with negative powers of $\varepsilon$ in its expression
via renormalized quantities form the quark-field renormalization constant $Z_q$.
The anomalous dimension is given by~(\ref{beta:gamma}):
\begin{equation*}
\gamma = - \frac{\beta}{3\beta_0}
\frac{N(-\beta,0)}{B(2+\beta,2+\beta) \Gamma(3+\beta) \Gamma(1-\beta)}
+\mathcal{O}\left(\frac{1}{\beta_0^2}\right)\,.
\end{equation*}
In the general covariant gauge, the one-loop term
proportional to $a$ should be added:
\begin{equation}
\begin{split}
\gamma_q &{}= C_F \frac{\alpha_s}{4\pi} \left[ 2a +
\frac{\beta \left(1 + \frac{2}{3}\beta\right)}
{B(2+\beta,2+\beta) \Gamma(3+\beta) \Gamma(1-\beta)} \right]
+\mathcal{O}\left(\frac{1}{\beta_0^2}\right)\\
&{}= C_F \frac{\alpha_s}{4\pi} \left[ 2a +
3 \beta \left(1 + \tfrac{5}{6}\beta - \tfrac{35}{36}\beta^2 + \cdots\right)
\right]\,.
\end{split}
\label{RL:gammaq}
\end{equation}
This perturbative series for $\gamma_q$ has the radius of convergence
equal to the distance from the origin to the nearest singularity,
which is situated at $\beta=-5/2$;
in other words, it converges at $|\beta|<5/2$.
It reproduces the leading-$\beta_0$ terms
in the 2-, 3-, 4-loop results~\cite{LV:93,CR:00}.

\begin{figure}[ht]
\begin{center}
\begin{picture}(116,38)
\put(32,12){\makebox(0,0){\includegraphics{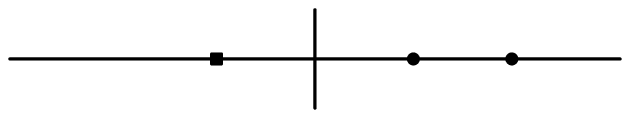}}}
\put(42,8.5){\makebox(0,0)[b]{$1$}}
\put(52,8.5){\makebox(0,0)[b]{$2$}}
\put(62,8.5){\makebox(0,0)[b]{$3$}}
\put(22,8.5){\makebox(0,0)[b]{$-1$}}
\put(12,8.5){\makebox(0,0)[b]{$-2$}}
\put(2,8.5){\makebox(0,0)[b]{$-3$}}
\put(32,0){\makebox(0,0)[b]{a}}
\put(95,22){\makebox(0,0){\includegraphics{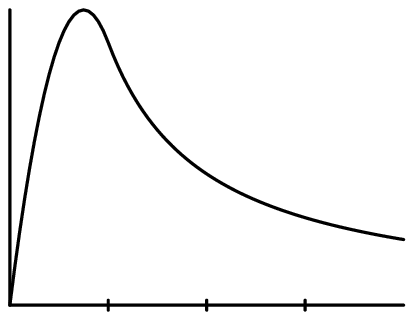}}}
\put(75,3.5){\makebox(0,0)[b]{$0$}}
\put(85,3.5){\makebox(0,0)[b]{$1$}}
\put(95,3.5){\makebox(0,0)[b]{$2$}}
\put(105,3.5){\makebox(0,0)[b]{$3$}}
\put(95,0){\makebox(0,0)[b]{b}}
\end{picture}
\end{center}
\caption{UV renormalons (black squares) and IR renormalons (black circles)
in the light-quark self-energy (a);
the virtuality distribution function (b)}
\label{Ren:Light}
\end{figure}

The renormalized expression for $\rlap/p S(p)$ is given by~(\ref{beta:A}).
If we factor out its $\mu$-dependence as in~(\ref{beta:RGsol}),
then the corresponding renormalization-group invariant
is given by~(\ref{beta:Ahat}) with
\begin{equation}
S(u) = \frac{1}{u} \left[\frac{N(0,u)}{(1+u)(1-u)(2-u)} - \frac{N(0,0)}{2}\right]
= - \frac{3C_F}{(1+u)(1-u)(2-u)}
\label{RL:S}
\end{equation}
(here $\sqrt{-p^2}$ plays the role of $m$).
The pole at $u=-1$ comes from the first $\Gamma$-function
in the numerator of~(\ref{RL:F}), and is an UV renormalon;
those at $u=1$, 2 come from the second $\Gamma$-function,
and are IR renormalons (Fig.~\ref{Ren:Light}a).
We can also see this from the power counting (Sect.~\ref{Sec:Ren}).
The light-quark self-energy seems to have a linear UV divergence.
However, the leading term of the integrand at $k\to\infty$,
$\rlap/k/(k^2)^2$, yields 0 after integration,
due to the Lorentz invariance.
The actual UV divergence is logarithmic: $\nu_{\text{UV}}=0$,
and UV renormalons can only be at $u\le0$.
The UV divergence at $u=0$ is removed by renormalization,
and UV renormalons are at $u<0$.
The index of the IR divergence of the self-energy,
like that of any off-shell quantity, is $\nu_{\text{IR}}=-2$,
and IR renormalons are at $u\ge1$.
Power corrections to the light-quark propagator form an expansion
in $1/(-p^2)$, therefore, IR renormalons can only appear
at positive integer values of $u$.
For gauge-invariant quantities, the first power correction
contains the gluon condensate ${<}G^2{>}$ of dimension 4,
and the first IR renormalon is at $u=2$.
The quark propagator is not gauge-invariant,
and the renormalon at $u=1$ is allowed.
The virtuality distribution function~(\ref{Ren:w}) is
\begin{equation*}
w(\tau) = - 3 C_F \times
\begin{cases}
\tfrac{1}{2}\tau - \tfrac{1}{3}\tau^2\,, & \tau < 1 \\
\tfrac{1}{6}\tau^{-1}\,,                 & \tau > 1
\end{cases}
\end{equation*}
(Fig.~\ref{Ren:Light}b).

\begin{figure}[ht]
\begin{center}
\begin{picture}(56,20)
\put(28,11.5){\makebox(0,0){\includegraphics{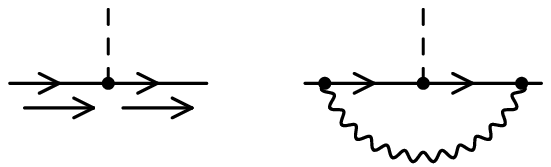}}}
\put(6,5){\makebox(0,0)[b]{$p$}}
\put(16,5){\makebox(0,0)[b]{$p'$}}
\put(11,0){\makebox(0,0)[b]{a}}
\put(43,0){\makebox(0,0)[b]{b}}
\end{picture}
\end{center}
\caption{Proper vertex $\Gamma(p,p')$ of a bilinear quark QCD current}
\label{HL:QCDvert}
\end{figure}

Now we shall discuss light-quark currents
\begin{equation*}
j_n(\mu) =  Z_{jn}^{-1}(\mu) \bar{q}'_0 \Gamma q_0\,,\quad
\Gamma = \gamma^{[\mu_1}\cdots\gamma^{\mu_n]}\,.
\end{equation*}
Let's calculate their vertex function $\Gamma(p,0)$
up to one loop (Fig.~\ref{HL:QCDvert}).
It is convenient to rewrite $\Gamma$ as
\begin{equation*}
\Gamma = \Gamma_+ + \Gamma_-\,,\quad
\Gamma_\pm = \frac{1}{2}
\left( \Gamma \pm \frac{\rlap/p\Gamma\rlap/p}{p^2} \right)\,,
\end{equation*}
where $\rlap/p\Gamma_\sigma=\sigma\Gamma_\sigma\rlap/p$, $\sigma=\pm1$.
Then
\begin{equation}
\gamma_\mu \Gamma_\sigma \gamma^\mu = 2 \sigma h(d) \Gamma_\sigma\,,
\label{RL:hdef}
\end{equation}
where for $n$ antisymmetrized $\gamma$-matrices
\begin{equation}
h(d) = \eta \left( n - \frac{d}{2} \right)\,,\quad
\eta = (-1)^{n+1} \sigma\,.
\label{RL:hres}
\end{equation}
The vertex function $\Gamma(p,0)$ for a Dirac matrix $\Gamma_\pm$
is $\Gamma_\pm\cdot\Gamma(p^2)$,
where the scalar function $\Gamma(p^2)$ can be calculated via $h(d)$,
once and for all Dirac matrices (see~\cite{BG:95}).

Calculating the vertex function $\Gamma(p^2)$ in the Landau gauge
with the denominator of the gluon propagator
raised to the power $n=1+(L-1)\varepsilon$,
we obtain~(\ref{beta:formF}), (\ref{RL:F}) with
\begin{equation}
N(\varepsilon,u) = - C_F \left[ 2-u-\varepsilon + 2h (u-h) \right]\,.
\label{RL:Nb}
\end{equation}
For the longitudinal vector current ($h=1-d/2$),
the result can be obtained from by differentiating $\Sigma(p^2)$
(Ward identity).
The anomalous dimension of the current is
\begin{equation*}
\gamma_{jn} = \frac{d\log Z_{\Gamma n}}{d\log\mu} + \gamma_q\,,
\end{equation*}
where the derivative of $Z_{\Gamma n}$ is given by~(\ref{beta:gamma}).
We arrive at~\cite{BG:95}
\begin{equation}
\begin{split}
\gamma_{jn} &{}= \frac{2}{3} C_F \frac{\alpha_s}{4\pi}
\frac{(n-1)(3-n+2\beta)}{B(2+\beta,2+\beta)\Gamma(3+\beta)\Gamma(1-\beta)}
+\mathcal{O}\left(\frac{1}{\beta_0^2}\right)\\
&{}= - 2 C_F \frac{\alpha_s}{4\pi} (n-1)
\left[ n-3 + \frac{n-15}{6} \beta - \frac{13n-35}{12} \beta^2 + \cdots \right]\,.
\end{split}
\label{RL:gammaj}
\end{equation}
Naturally, it vanishes at $n=1$.
This perturbative series converges at $|\beta|<5/2$.
It reproduces the leading powers of $\beta_0$
in the two- and three-loop results~\cite{BG:95,Gr:00}
In particular, the mass anomalous dimension is
\begin{equation}
\gamma_m = - \gamma_{j0} = 2 C_F \frac{\alpha_s}{4\pi}
\frac{1+\frac{2}{3}\beta}{B(2+\beta,2+\beta) \Gamma(3+\beta) \Gamma(1-\beta)}
+\mathcal{O}\left(\frac{1}{\beta_0^2}\right)
\label{RL:gammam}
\end{equation}
(it is known at four loops~\cite{Ch:97,VLR:97}).

There is a general belief that one may use
a naively anticommutating $\gamma_5^{\text{AC}}$ in open quark lines
without encountering contradictions, see~\cite{La:93}.
The pseudoscalar currents with $\gamma_5^{\text{AC}}$,
$j_{\text{AC}}(\mu)=
Z_{P,\text{AC}}^{-1}(\mu)\bar{q}'_0\gamma_5^{\text{AC}}q$,
and with the 't~Hooft--Veltman $\gamma_5^{\text{HV}}$,
$j_{\text{HV}}(\mu)=
Z_{P,\text{HV}}^{-1}(\mu)\bar{q}'_0\gamma_5^{\text{HV}}q$
are related to each other by a finite renormalization
\begin{equation}
j_{\text{AC}}(\mu) = Z_P(\alpha_s(\mu)) j_{\text{HV}}(\mu)\,,\quad
Z_P(\alpha_s) = 1 + z_{\text{P}1} \frac{\alpha_s}{4\pi}
+ z_{\text{P}2} \left(\frac{\alpha_s}{4\pi}\right)^2 + \cdots
\label{g5:ZPdef}
\end{equation}
Similarly, the axial currents are related by
\begin{equation}
j_{\text{AC}}^\mu(\mu) = Z_A(\alpha_s(\mu)) j_{\text{HV}}^\mu(\mu)\,.
\label{g5:ZAdef}
\end{equation}
The finite renormalization constants $Z_{P,A}(\alpha_s)$
can be obtained from the currents' anomalous dimensions.
Multiplying the Dirac matrix $\Gamma$ in the current by $\gamma_5^{\text{AC}}$
does not change the anomalous dimension;
multiplying by $\gamma_5^{\text{HV}}$ means $n\to4-n$.
Differentiating~(\ref{g5:ZPdef}) and~(\ref{g5:ZAdef}), we have
\begin{equation}
\begin{split}
\frac{d\log Z_{\text{P}}(\alpha_s)}{d\log\alpha_s} &{}=
\frac{\gamma_{j0}(\alpha_s)-\gamma_{j4}(\alpha_s)}{2\beta(\alpha_s)}\,,\\
\frac{d\log Z_{\text{A}}(\alpha_s)}{d\log\alpha_s} &{}=
\frac{\gamma_{j1}(\alpha_s)-\gamma_{j3}(\alpha_s)}{2\beta(\alpha_s)}
\quad\text{where {}}\gamma_{j1}=0\,.
\end{split}
\label{g5:ZPA}
\end{equation}
Therefore,
\begin{equation}
Z_P(\alpha_s) = K_{\gamma_{j0}-\gamma_{j4}}(\alpha_s)\,,\quad
Z_A(\alpha_s) = K_{\gamma_{j1}-\gamma_{j3}}(\alpha_s)
\label{g5:K}
\end{equation}
(see~(\ref{CMag:K})).
For the tensor current, multiplying $\sigma^{\mu\nu}$ by $\gamma_5^{\text{HV}}$
is merely a space-time transformation,
e.g., $\gamma_5^{\text{HV}}\sigma^{01}=-i\sigma^{23}$,
and thus it does not change the anomalous dimension.
Therefore, the constant relating the currents
$\bar{q}'\gamma_5^{\text{AC}}\sigma^{\mu\nu}q$ and
$\bar{q}'\gamma_5^{\text{HV}}\sigma^{\mu\nu}q$ is
\begin{equation}
Z_T(\alpha_s) = 1\,.
\label{g5:ZT}
\end{equation}

In the large-$\beta_0$ limit, we obtain from~(\ref{g5:ZPA}), (\ref{RL:gammaj})
\begin{equation}
\begin{split}
Z_A &{}= 1 - \frac{4}{3} \frac{C_F}{\beta_0} \int_0^\beta
\frac{d\beta}{B(2+\beta,2+\beta)\Gamma(3+\beta)\Gamma(1-\beta)}\\
&{}= 1 - 4 C_F \frac{\alpha_s}{4\pi}
\left[ 1 + \frac{1}{12} \beta - \frac{13}{36} \beta^2 + \cdots \right]\,,\\
Z_{\text{P}} &{}= Z_{\text{A}}^2\,.
\end{split}
\label{RL:ZPA}
\end{equation}
This reproduces the leading powers of $\beta_0$
in the three-loop results~\cite{La:93}.

\section{Heavy Quark in HQET}
\label{Sec:RH}

\begin{figure}[ht]
\begin{center}
\begin{picture}(32,9.5)
\put(16,4.75){\makebox(0,0){\includegraphics{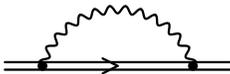}}}
\end{picture}
\end{center}
\caption{One-loop heavy quark self-energy in HQET}
\label{Renorm:HQET1Loop}
\end{figure}

Now we turn to Heavy Quark Effective Theory
(HQET, see, e.g., \cite{Ne:94,MW:00,G:03})
and discuss the heavy-quark propagator.
The one-loop expression for the self-energy $\widetilde{\Sigma}(\omega)/\omega$
(Fig.~\ref{Renorm:HQET1Loop}) in the Landau gauge
with the gluon denominator raised to the power $n=1+(L-1)\varepsilon$ is
\begin{equation*}
a_1(n) = \frac{iC_F}{\omega^2} \int \frac{d^d k}{\pi^{d/2}}
\frac{\omega}{k\cdot v+\omega} \frac{v^\mu v^\nu}{(-k^2)^n}
\left(g_{\mu\nu} + \frac{k_\mu k_\nu}{-k^2}\right)\,.
\end{equation*}
Using the one-loop HQET integrals
\begin{equation}
\begin{split}
&\int \frac{d^d k}{(-k^2)^{n_2}}
\left( \frac{\omega}{k\cdot v+\omega}\right)^{n_1} =
i \pi^{d/2} (-2\omega)^{d-2n_2} I(n_1,n_2)\,,\\
&I(n_1,n_2) =
\frac{\Gamma(-d+n_1+2n_2) \Gamma(d/2-n_2)}{\Gamma(n_1) \Gamma(n_2)}\,,
\end{split}
\label{HQET1:I1}
\end{equation}
we can easily find the function $F(\varepsilon,u)$ (\ref{beta:F}).
Such functions for all off-shell HQET quantities have
the same $\Gamma$-function structure resulting from~(\ref{HQET1:I1})
with $n_2=1+u-\varepsilon$:
\begin{equation}
F(\varepsilon,u) = \left(\frac{\mu}{-2\omega}\right)^{2u} e^{\gamma\varepsilon}
\frac{u \Gamma(-1+2u) \Gamma(1-u)}{\Gamma(2+u-\varepsilon)}
D(\varepsilon)^{u/\varepsilon-1} N(\varepsilon,u)\,.
\label{RH:F}
\end{equation}
The first $\Gamma$-function in the numerator,
with the positive sign in front of $u$,
comes from the first $\Gamma$-function in the numerator of~(\ref{HQET1:I1}),
with the negative sign in front of $d$,
and its poles are UV divergences.
The second $\Gamma$-function in the numerator,
with the negative sign in front of $u$,
comes from the second $\Gamma$-function in the numerator of~(\ref{HQET1:I1}),
with the positive sign in front of $d$,
and its poles are IR divergences.
For $\widetilde{\Sigma}(\omega)/\omega$, we obtain
\begin{equation}
N(\varepsilon,u) = - 2 C_F (3-2\varepsilon)\,.
\label{RH:N}
\end{equation}
If $a_0\ne0$, the one-loop term proportional to $a_0$
should be added to $\widetilde{\Sigma}(\omega)$.
This formula reproduces the largest powers of $\beta_0$
in the known results~\cite{BG:91}
at $L=u/\varepsilon=1$ and $L=2$.

The heavy-quark propagator
$\widetilde{S}(\omega)=\bigl[\omega-\widetilde{\Sigma}(\omega)\bigr]^{-1}$
with the $1/\beta_0$ accuracy in the Landau gauge
is equal to $1/\omega$ times~(\ref{beta:formF}),
where $F(\varepsilon,u)$ is given by~(\ref{RH:F}), (\ref{RH:N}).
Terms with negative powers of $\varepsilon$ in its expression
via renormalized quantities form $\widetilde{Z}_Q$.
The anomalous dimension is given by~(\ref{beta:gamma}):
\begin{equation*}
\widetilde{\gamma} = \frac{\beta}{6\beta_0}
\frac{N(-\beta,0)}{B(2+\beta,2+\beta) \Gamma(2+\beta) \Gamma(1-\beta)}
+\mathcal{O}\left(\frac{1}{\beta_0^2}\right)\,.
\end{equation*}
In the general covariant gauge, the one-loop term
proportional to $a$ should be added:
\begin{equation}
\begin{split}
\widetilde{\gamma}_Q &{}= C_F \frac{\alpha_s}{4\pi} \left[ 2a -
\frac{1 + \frac{2}{3}\beta}
{B(2+\beta,2+\beta) \Gamma(2+\beta) \Gamma(1-\beta)} \right]
+\mathcal{O}\left(\frac{1}{\beta_0^2}\right)\\
&{}= C_F \frac{\alpha_s}{4\pi} \left[ 2(a-3) - 8\beta + \tfrac{10}{3}\beta^2 + \cdots
\right]\,.
\end{split}
\label{RH:gammaQ}
\end{equation}
This perturbative series converges at $|\beta|<5/2$.
It reproduces the leading-$\beta_0$ terms
in the two- and three-loop results~\cite{BG:91,MR:00,CG:03}.

\begin{figure}[ht]
\begin{center}
\begin{picture}(84,35)
\put(42,29){\makebox(0,0){\includegraphics{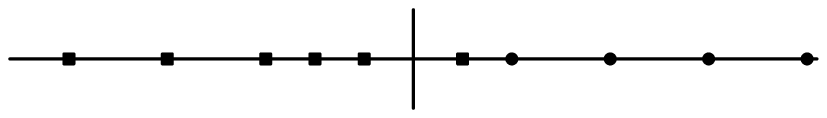}}}
\put(52,25.5){\makebox(0,0)[b]{$1$}}
\put(62,25.5){\makebox(0,0)[b]{$2$}}
\put(72,25.5){\makebox(0,0)[b]{$3$}}
\put(82,25.5){\makebox(0,0)[b]{$4$}}
\put(32,25.5){\makebox(0,0)[b]{$-1$}}
\put(22,25.5){\makebox(0,0)[b]{$-2$}}
\put(12,25.5){\makebox(0,0)[b]{$-3$}}
\put(2,25.5){\makebox(0,0)[b]{$-4$}}
\put(42,20){\makebox(0,0)[b]{a}}
\put(42,9){\makebox(0,0){\includegraphics{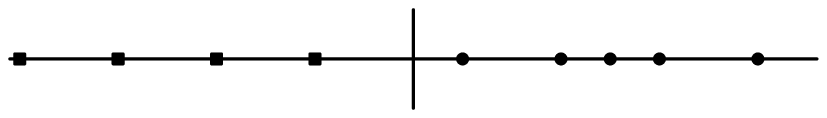}}}
\put(52,5.5){\makebox(0,0)[b]{$1$}}
\put(62,5.5){\makebox(0,0)[b]{$2$}}
\put(72,5.5){\makebox(0,0)[b]{$3$}}
\put(82,5.5){\makebox(0,0)[b]{$4$}}
\put(32,5.5){\makebox(0,0)[b]{$-1$}}
\put(22,5.5){\makebox(0,0)[b]{$-2$}}
\put(12,5.5){\makebox(0,0)[b]{$-3$}}
\put(2,5.5){\makebox(0,0)[b]{$-4$}}
\put(42,0){\makebox(0,0)[b]{b}}
\end{picture}
\end{center}
\caption{Renormalons in the off-shell HQET self-energy (a)
and on-shell heavy-quark self-energy (b)}
\label{Ren:Heavy}
\end{figure}

The renormalized expression for $\omega\widetilde{S}(\omega)$ is given by~(\ref{beta:A}).
If we factor out its $\mu$-dependence as in~(\ref{beta:RGsol}),
then the corresponding renormalization-group invariant
is given by~(\ref{beta:Ahat}) with~\cite{BB:94}
\begin{equation}
S(u) = \frac{\Gamma(-1+2u) \Gamma(1-u)}{\Gamma(2+u)} N(0,u) + \frac{N(0,0)}{2u}
= - 6 C_F \left[ \frac{\Gamma(-1+2u) \Gamma(1-u)}{\Gamma(2+u)} + \frac{1}{2u} \right]
\label{RH:S}
\end{equation}
(here $-2\omega$ plays the role of $m$).
The first $\Gamma$-function, with the positive sign in front of $u$,
produces UV renormalons,
while the second one, with the negative sign,
produces IR renormalons (Fig.~\ref{Ren:Heavy}a).
We can understand this from the power counting (Sect.~\ref{Sec:Ren}).
The heavy-quark self-energy has a linear UV divergence
which is not nullified by the Lorentz invariance: $\nu_{\text{UV}}=1$.
This is the same divergence as that of the Coulomb energy
of a point charge in classical electrodynamics.
Therefore, UV renormalons are situated at $u\le1/2$.
The index of the IR divergence of the self-energy,
like that of any off-shell quantity, is $\nu_{\text{IR}}=-2$,
and IR renormalons are at $u\ge1$.

Here we encounter a radically new situation: an UV renormalon at $u>0$.
It leads to the ambiguity
$\Delta\widetilde{\Sigma}(\omega)/\omega=(r/\beta_0)e^{5/5}\Lambda_{\MS}/(-2\omega)$
where $r=4C_F$ is the residue of $S(u)$ at $u=1/2$.
If we change the prescription for handling the pole at $u=1/2$,
we have to change the zero-energy level of HQET.
Therefore, the HQET meson energy has an ambiguity
$\Delta\bar{\Lambda}=\Delta\widetilde{\Sigma}(\omega)$
of order $\Lambda_{\MS}/\beta_0$~\cite{BB:94}
(see also~\cite{BSUV:94})
\begin{equation}
\Delta\bar{\Lambda} = - 2 C_F e^{5/6} \frac{\Lambda_{\MS}}{\beta_0}\,.
\label{RH:DL}
\end{equation}

\begin{figure}[ht]
\begin{center}
\begin{picture}(42,22)
\put(21,11){\makebox(0,0){\includegraphics{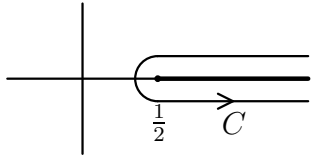}}}
\put(21,5){\makebox(0,0){$\frac{1}{2}$}}
\put(31,5){\makebox(0,0){$C$}}
\end{picture}
\end{center}
\caption{Integration contour}
\label{Ren:Cut}
\end{figure}

The structure of the leading UV renormalon at $u=1/2$
can be investigated beyond the large-$\beta_0$ limit~\cite{Be:95}.
This approach is based on the renormalization group~\cite{Pa:78,BBK:97}.
The renormalization-group invariant corresponding to $\omega\widetilde{S}(\omega)$
is now written as
\begin{equation}
1 + \frac{1}{\beta_0} \int_0^\infty d u\, S(u)\,
\exp\left[-\frac{4\pi}{\beta_0\alpha_s(\mu_0)} u\right]
\label{RH:exact}
\end{equation}
instead of~(\ref{beta:Ahat}),
where the exact $\alpha_s$ is used in the exponent,
$\mu_0=-2\omega e^{-5/6}$,
and $\mathcal{O}(1/\beta_0^2)$ is absent.
The singularity of $S(u)$ at $u=1/2$ becomes a branching point
\begin{equation*}
S(u) = \frac{r}{\left(\frac{1}{2}-u\right)^{1+a}}+\cdots\,,
\end{equation*}
with the cut from $1/2$ to $+\infty$,
instead of the simple pole.
The renormalon ambiguity of $\widetilde{\Sigma}(\omega)/\omega$ is defined,
as before, as the difference of the integrals~(\ref{RH:exact})
below and above the real axis divided by $2\pi i$:
\begin{equation}
\begin{split}
\Delta\bar{\Lambda} &{}= \frac{r\omega}{\beta_0} \frac{1}{2\pi i} \int_C
\frac{d u}{\left(\frac{1}{2}-u\right)^{1+a}}
\exp\left[-\frac{4\pi}{\beta_0\alpha_s(\mu_0)} u\right]\\
&{}= \frac{r}{2\beta_0\Gamma(1+a)} (-2\omega)
\exp\left[-\frac{2\pi}{\beta_0\alpha_s(\mu_0)}\right]
\left(\frac{\beta_0\alpha_s(\mu_0)}{4\pi}\right)^{-a}
\end{split}
\label{RH:D}
\end{equation}
(Fig.~\ref{Ren:Cut}; we have used $\Gamma(-a)\Gamma(1+a)=-\pi/\sin\pi a$).
But this must be just some number times $\Lambda_{\MS}$,
and cannot depend on $\omega$!

We have to use a formula for $\alpha_s(\mu)$ more precise
than the one-loop one~(\ref{beta:as}).
The renormalization-group equation
is solved by separation of variables:
\begin{gather*}
\frac{2\pi}{\beta_0} \int \frac{d\alpha_s}{\alpha_s^2} \left[ 1
- \frac{\beta_1}{\beta_0} \frac{\alpha_s}{4\pi} + \mathcal{O}(\alpha_s^2) \right]
= - \int d\log\mu\,,\\
\frac{2\pi}{\beta_0\alpha_s(\mu)}
+ \frac{\beta_1}{2\beta_0^2} \log \frac{\alpha_s(\mu)}{4\pi}
+ \mathcal{O}(\alpha_s) = \log\frac{\mu}{\Lambda_{\MS}}\,,
\end{gather*}
and hence
\begin{equation}
\Lambda_{\MS} = \mu \exp\left[-\frac{2\pi}{\beta_0\alpha_s(\mu)}\right]
\left(\frac{\alpha_s(\mu)}{4\pi}\right)^{-\frac{\beta_1}{2\beta_0^2}}
\left[1+\mathcal{O}(\alpha_s)\right]\,.
\label{RH:Lambda}
\end{equation}

The UV renormalon ambiguity $\Delta\bar{\Lambda}$ must be equal to
$\Lambda_{\MS}$ times some number:
\begin{equation}
\Delta \bar{\Lambda} = N_0 \Delta_0\,,\quad
\Delta_0 = - 2 C_F e^{5/6} \frac{\Lambda_{\MS}}{\beta_0}\,.
\label{RH:DLgen}
\end{equation}
The normalization factor $N_0$ is only known in the large-$\beta_0$ limit:
\begin{equation*}
N_0 = 1 + \mathcal{O}(1/\beta_0)\,;
\end{equation*}
in general, it is just some unknown number of order 1.
Comparing~(\ref{RH:D}) with~(\ref{RH:DLgen}),
we conclude that at $u\to1/2$
\begin{equation}
S(u) = - \frac{4 C_F N_0'}{\left(\frac{1}{2}-u\right)^{1+\frac{\beta_1}{2\beta_0^2}}}
\left[1 + \mathcal{O}\left(\tfrac{1}{2}-u\right)\right]\,,
\label{RH:Branch}
\end{equation}
where
\begin{equation}
N_0' = N_0 \Gamma\left(1+\frac{\beta_1}{2\beta_0^2}\right)
\beta_0^{\frac{\beta_1}{2\beta_0^2}}\,.
\label{RH:N0p}
\end{equation}
The result for the power is exact;
the normalization cannot be found within this approach.

\begin{figure}[ht]
\begin{center}
\begin{picture}(69,10)
\put(34.5,5){\makebox(0,0){\includegraphics{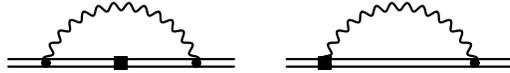}}}
\end{picture}
\end{center}
\caption{One-loop diagrams for $\widetilde{\Sigma}_{k}$}
\label{Mcor:Sk0}
\end{figure}

The self-energy with a kinetic-energy insertion $\widetilde{\Sigma}_{k}$
(Fig.~\ref{Mcor:Sk0})
can be also easily calculated in the large-$\beta_0$ limit.
In the Landau gauge, raising the gluon denominator
to the power $n=1+(L-1)\varepsilon$:
\begin{equation*}
\widetilde{\Sigma}_{k}(\omega) = i C_F g_0^2 \int \frac{d^d k}{(2\pi)^d}
\frac{v^\mu\left[2(k\cdot v+\omega)k_\bot^\nu-k_\bot^2 v^\nu\right]}
{(-k^2)^n(k\cdot v+\omega)^2}
\left[g_{\mu\nu} + \frac{k_\mu k_\nu}{-k^2}\right]\,,
\end{equation*}
we obtain~(\ref{RH:F}) with
\begin{equation}
N(\varepsilon,u) = 2 C_F (3-2\varepsilon)^2 \omega^2\,,
\label{RH:Sk0}
\end{equation}
and hence
\begin{equation}
\Delta \widetilde{\Sigma}_{k}(\omega) = - 3 \omega \Delta \bar{\Lambda}\,.
\label{RH:DSk0}
\end{equation}
This leads to the UV renormalon ambiguity
of the heavy-quark field renormalization constant~\cite{GN:97}
\begin{equation}
\Delta \widetilde{Z}_Q = - \frac{3}{2} \frac{\Delta \bar{\Lambda}}{m}\,.
\label{RH:DZQ}
\end{equation}

\begin{figure}[ht]
\begin{center}
\begin{picture}(56,20)
\put(28,11.5){\makebox(0,0){\includegraphics{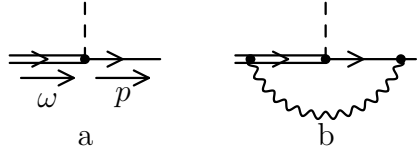}}}
\put(6,5){\makebox(0,0)[b]{$\omega$}}
\put(16,5){\makebox(0,0)[b]{$p$}}
\put(11,0){\makebox(0,0)[b]{a}}
\put(43,0){\makebox(0,0)[b]{b}}
\end{picture}
\end{center}
\caption{Proper vertex $\widetilde{\Gamma}(\omega,p)$ of a heavy--light HQET current}
\label{HL:HQETvert}
\end{figure}

Let's now discuss the heavy--light quark current in HQET.
If the light quark is massless,
we may take $\frac{1}{4}\Tr$ of $\gamma$-matrices on the light-quark line
of any diagram for $\widetilde{\Gamma}(\omega,0)$.
All diagrams with insertions to the gluon propagator
of the one-loop diagram (Fig.~\ref{HL:HQETvert}b),
as well as this one-loop diagram itself in the Landau gauge,
vanish due to transversality of the gluon propagator.
Therefore, to the first order in $1/\beta_0$, $\widetilde{\Gamma}(\omega,0)=1$,
and $\widetilde{\gamma}_j=\frac{1}{2}(\widetilde{\gamma}_Q+\gamma_q)$ in the Landau gauge.
This anomalous dimension is gauge-invariant, and~\cite{BG:95}
\begin{equation}
\begin{split}
\widetilde{\gamma}_j &{}= - C_F \frac{\alpha_s}{4\pi}
\frac{1+\frac{2}{3}\beta}{B(2+\beta,2+\beta)\Gamma(3+\beta)\Gamma(1-\beta)}
+ \mathcal{O}\left(\frac{1}{\beta_0^2}\right)\\
&{}= - 3 C_F \frac{\alpha_s}{4\pi} \left[ 1 + \tfrac{5}{6} \beta
- \tfrac{35}{36} \beta^2 + \cdots \right]\,,
\end{split}
\label{RHL:gammaj}
\end{equation}
from~(\ref{RL:gammaq}) and~(\ref{RH:gammaQ}).
This perturbative series converges at $\beta_0|\alpha_s|<4\pi$.
It reproduces the leading-$\beta_0$ terms
in the two- and three-loop results~\cite{BG:91,JM:91,CG:03}.
Note that $\widetilde{\gamma}_j=\frac{1}{2}\gamma_{j0}$~(\ref{RL:gammaj})
at the first order in $1/\beta_0$.

\begin{figure}[ht]
\begin{center}
\begin{picture}(42,25)
\put(21,11){\makebox(0,0){\includegraphics{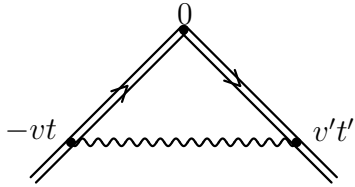}}}
\put(21,23){\makebox(0,0){0}}
\put(38,6){\makebox(0,0)[bl]{$v't'$}}
\put(4,6){\makebox(0,0)[br]{$-vt$}}
\end{picture}
\end{center}
\caption{One-loop heavy--heavy vertex}
\label{HH:Vert1}
\end{figure}

Finally, we discuss the heavy--heavy current
\begin{equation*}
\widetilde{J} = \widetilde{Z}_J^{-1}(\cosh\vartheta) \widetilde{J}_0\,,\quad
\widetilde{J}_0 = \widetilde{Q}^+_{v'0}\widetilde{Q}_{v0}\,,\quad
\cosh\vartheta = v\cdot v'
\end{equation*}
in HQET.
At one loop (Fig.~\ref{HH:Vert1}), we use the Fourier transform
of the Landau-gauge gluon propagator with the denominator raised to the power $n$,
\begin{equation*}
\frac{i}{2^{2n-1}\pi^{d/2}}\,\frac{\Gamma(d/2-n)}{\Gamma(n+1)}\,
\frac{(2n-1) x^2 g_{\mu\nu} + (d-2n) x_\mu x_\nu}{(-x^2+i0)^{d/2-n+1}}\,.
\end{equation*}
to obtain the coordinate-space vertex
\begin{equation*}
\begin{split}
&\widetilde{\Gamma}_1(t,t';\cosh\vartheta) =
- C_F \frac{g_0^2}{2^{2n+1}\pi^{d/2}} \frac{\Gamma(d/2-n)}{\Gamma(n+1)}
\theta(t) \theta(t')\\
&{}\times
\frac{(2n-1)x^2\cosh\vartheta+(d-2n)(t+t'\cosh\vartheta)(t'+t\cosh\vartheta)}%
{(-x^2)^{d/2}}\,,\\
&x^2=t^2+t^{\prime2}+2tt'\cosh\vartheta\,.
\end{split}
\end{equation*}

The momentum-space vertex function is expressed via the coordinate-space one as
\begin{equation*}
\widetilde{\Gamma}(\omega,\omega';\cosh\vartheta) =
\int dt\,dt'\,e^{i\omega t+i\omega't'}
\widetilde{\Gamma}(t,t';\cosh\vartheta)\,.
\end{equation*}
Ultraviolet divergences of $\widetilde{\Gamma}(\omega,\omega';\cosh\vartheta)$
do not depend on the residual energies $\omega$, $\omega'$, and we may nullify them.
An infrared cutoff is then necessary to avoid IR $1/\varepsilon$ terms.
Proceeding to the variables
\begin{equation*}
t = \tau \frac{1+\xi}{2}\,,\quad t' = \tau \frac{1-\xi}{2}\,,
\end{equation*}
we obtain the coefficient $a_1(n)$
\begin{equation*}
\begin{split}
&a_1(n) = - C_F 2^{d-2n-2} \frac{\Gamma(d/2-n)}{\Gamma(n+1)}
\int_0^T \frac{d\tau}{\tau^{d-2n-1}} \times{}\\
&\int_{-1}^{+1} d\xi\,
\frac{(2n-1) \cosh\vartheta (c^2-s^2\xi^2) + (d-2n) (c^4-s^4\xi^2)}%
{(-c^2+s^2\xi^2)^{d/2-n+1}}\,,
\end{split}
\end{equation*}
where $c=\cosh\frac{\vartheta}{2}$, $s=\sinh\frac{\vartheta}{2}$,
and the upper limit $T$ provides an infrared cutoff.
Changing the integration variable $\xi=\tanh\psi/\tanh(\vartheta/2)$,
we obtain
\begin{equation*}
\begin{split}
a_1(n) ={}& C_F \frac{\Gamma(d/2-n-1)}{\Gamma(n+1)}
\left(\frac{i}{2} T \cosh\frac{\vartheta}{2}\right)^{2n+2-d}
\int_{-\vartheta/2}^{+\vartheta/2} \frac{d\psi}{\cosh^{2n+2-d}\psi}\\
&{}\times\left[ \left(\frac{d}{2}+n-1\right) \coth\vartheta
+ \frac{d/2-n}{\sinh\vartheta} \cosh 2\psi \right]
\end{split}
\end{equation*}
(it becomes real in the Euclidean space $T\to-i T_{E}$).
Therefore~(\ref{beta:F}),
\begin{equation*}
\begin{split}
F(\varepsilon,u) ={}& - C_F \frac{\Gamma(1-u)}{\Gamma(2+u-\varepsilon)}
e^{\gamma\varepsilon} D(\varepsilon)^{u/\varepsilon-1}
\left(\frac{i}{2} \mu T \cosh\frac{\vartheta}{2}\right)^{2u}\\
&{}\times\int_{-\vartheta/2}^{+\vartheta/2} \frac{d\psi}{\cosh^{2u}\psi}\,
\left[ (2+u-2\varepsilon) \coth\vartheta
+ \frac{1-u}{\sinh\vartheta} \sinh 2\psi \right]\,.
\end{split}
\end{equation*}
The anomalous dimension corresponding to $\widetilde{Z}_\Gamma$ is~(\ref{beta:gamma})
\begin{equation*}
\widetilde{\gamma}_\Gamma = \frac{1}{3} C_F \frac{\alpha_s}{4\pi}
\frac{2 (1+\beta) \vartheta \coth\vartheta + 1}%
{B(2+\beta,2+\beta) \Gamma(2+\beta) \Gamma(1-\beta)}\,.
\end{equation*}
In order to obtain $\widetilde{\gamma}_J=\widetilde{\gamma}_\Gamma+\widetilde{\gamma}_Q$,
we add~(\ref{RH:gammaQ}):
\begin{equation}
\begin{split}
\widetilde{\gamma}_J &{}= \frac{2}{3} C_F \frac{\alpha_s}{4\pi}
\frac{\vartheta \coth\vartheta - 1}%
{B(2+\beta,2+\beta) \Gamma(1+\beta) \Gamma(1-\beta)}\\
&{}= 4 C_F \frac{\alpha_s}{4\pi}
\left(1 + \frac{5}{3} \beta - \frac{1}{3} \beta^2 + \cdots\right)
\left(\vartheta \coth\vartheta - 1\right)\,.
\end{split}
\label{RH:gammaJ}
\end{equation}
It vanishes at $\vartheta=0$ as expected.
It reproduces the leading $\beta_0$ terms in the two-loop result~\cite{KR:87}.

\section{On-shell Heavy Quark in QCD}
\label{Sec:RHos}

\begin{figure}[ht]
\begin{center}
\begin{picture}(32,16)
\put(16,8){\makebox(0,0){\includegraphics{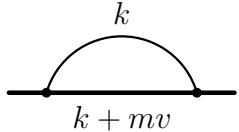}}}
\put(16,1){\makebox(0,0){$k+mv$}}
\put(16,15){\makebox(0,0){$k$}}
\end{picture}
\end{center}
\caption{One-loop on-shell heavy-quark self-energy}
\label{Mcor:OS1}
\end{figure}

Now, we turn to the on-shell mass and wave-function renormalization
of a heavy quark in QCD at the order $1/\beta_0$.
It is convenient~\cite{MR:00} to introduce the function
\begin{equation*}
T(t) = \frac{1}{4m} \Tr (\rlap/v+1) \Sigma(mv(1+t))\,,
\end{equation*}
then the renormalization constants are
\begin{equation*}
Z_m^{\text{os}}=1-T(0)\,,\quad
Z_Q^{\text{os}}=\left[1-T'(0)\right]^{-1}\,.
\end{equation*}
At one loop (Fig.~\ref{Mcor:OS1}),
in the Landau gauge,
with the gluon denominator raised to the power $n=1+(L-1)\varepsilon$,
we have
\begin{equation*}
\begin{split}
a_1(n) &{}= -i C_F \int \frac{d^d k}{\pi^{d/2}}
\frac{\Tr(\rlap/v+1)\gamma^\mu(\rlap/p+\rlap/k+m)\gamma^\nu}{4m D_1(t) D_2^n}
\left[g_{\mu\nu} + \frac{k_\mu k_\nu}{D_2}\right]\\
&{}= - 2 i C_F \int \frac{d^d k}{\pi^{d/2}} \frac{1}{D_1(t) D_2^n}
\left[1 - \frac{(d-2) D_2}{4 m^2} (1-t) + \mathcal{O}(t^2)\right]\,,
\end{split}
\end{equation*}
where
\begin{equation*}
p = mv(1+t)\,,\quad
D_1(t) = m^2 - (k+p)^2\,,\quad
D_2 = -k^2\,,
\end{equation*}
Expanding $1/D_1(t)$ to the linear term in $t$
and using the one-loop on-shell integrals
\begin{equation}
\begin{split}
&\int \frac{d^d k}{\left[m^2-(k+mv)^2\right]^{n_1}(-k^2)^{n_2}} =
i \pi^{d/2} m^{d-2(n_1+n_2)} M(n_1,n_2)\,,\\
&M(n_1,n_2) =
\frac{\Gamma(d-n_1-2n_2)\Gamma(-d/2+n_1+n_2)}{\Gamma(n_1)\Gamma(d-n_1-n_2)}\,,
\end{split}
\label{QCDos:M1}
\end{equation}
we find $F(\varepsilon,u)$~(\ref{beta:F}).

The functions $F$ for all on-shell quantities have a common
$\Gamma$-function structure resulting from~(\ref{QCDos:M1})
with $n_2=1+u-\varepsilon$:
\begin{equation}
F(\varepsilon,u) = \left(\frac{\mu}{m}\right)^{2u} e^{\gamma\varepsilon}
\frac{\Gamma(1+u) \Gamma(1-2u)}{\Gamma(3-u-\varepsilon)}
D(\varepsilon)^{u/\varepsilon-1} N(\varepsilon,u)\,.
\label{RHos:N}
\end{equation}
The first $\Gamma$-function in the numerator,
with the positive sign in front of $u$,
comes from the second $\Gamma$-function in the numerator of~(\ref{QCDos:M1}),
with the negative sign in front of $d$,
and its poles are UV divergences.
The second $\Gamma$-function in the numerator,
with the negative sign in front of $u$,
comes from the first $\Gamma$-function in the numerator of~(\ref{QCDos:M1}),
with the positive sign in front of $d$,
and its poles are IR divergences.
For $T(t)$, we obtain~\cite{BG:95}
\begin{equation}
N(\varepsilon,u) = 2 C_F (3-2\varepsilon) (1-u)
\bigl[1-(1+u-\varepsilon)t\bigr] + \mathcal{O}(t^2)\,.
\label{RHos:NT}
\end{equation}

The on-shell mass renormalization constant $Z_m^{\text{os}}=m_0/m=1-T(0)$
with the $1/\beta_0$ accuracy is given by~(\ref{beta:formF}), (\ref{RHos:N})
with $N(\varepsilon,u)$ equal to minus~(\ref{RHos:NT}) at $t=0$.
Retaining only terms with negative powers of $\varepsilon$,
we obtain the \MS{} mass renormalization constant $Z_m$
(because $Z_m^{\text{os}}$ contains no IR divergences).
Using~(\ref{beta:gamma}),
we reproduce the mass anomalous dimension~(\ref{RL:gammam}).
Retaining terms with $\varepsilon^0$,
we get $Z_m^{\text{os}}/Z_m(\mu)=m(\mu)/m$ in the form~(\ref{beta:A}).
As usual, it is convenient to express $m(\mu)$
via the renormalization-group invariant $\hat{m}$~(\ref{beta:RGsol}).
Then the ratio~\cite{BB:94}
\begin{equation}
\begin{split}
\frac{m}{\hat{m}} &{}= 1
+ \frac{1}{\beta_0} \int_0^\infty d u\,e^{-u/\beta} S(u)
+ \mathcal{O}\left(\frac{1}{\beta_0^2}\right)\,,\\
S(u) &{}= 6 C_F \left[ \frac{\Gamma(u) \Gamma(1-2u)}{\Gamma(3-u)} (1-u)
- \frac{1}{2u} \right]\,.
\end{split}
\label{RHos:mm}
\end{equation}
The first $\Gamma$-function, with the positive sign in front of $u$,
produces UV renormalons,
while the second one, with the negative sign,
produces IR renormalons (Fig.~\ref{Ren:Heavy}b).
We can understand this from the power counting (Sect.~\ref{Sec:Ren}).
The QCD quark self-energy has a logarithmic UV divergence
($\nu_{\text{UV}}=0$), and hence UV renormalons are situated at $u<0$.
The index of the IR divergence of the on-shell quark self-energy
is $\nu_{\text{IR}}=-1$, and IR renormalons are at $u\ge1/2$.

\begin{figure}[ht]
\begin{center}
\begin{picture}(42,33)
\put(21,16.5){\makebox(0,0){\includegraphics{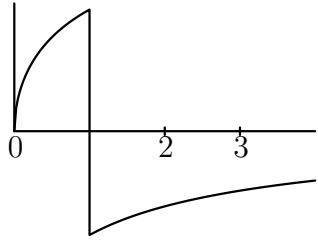}}}
\put(1,11.5){\makebox(0,0)[b]{$0$}}
\put(21,11.5){\makebox(0,0)[b]{$2$}}
\put(31,11.5){\makebox(0,0)[b]{$3$}}
\end{picture}
\end{center}
\caption{Virtuality distribution function}
\label{Ren:Mass}
\end{figure}

The ratio~(\ref{RHos:mm}) can be represented~\cite{Ne:95}
in the form~(\ref{Ren:Neubert}).
At $\tau>1$, the distribution function is given by the sum~(\ref{Ren:winf})
over the UV renormalons at $u=-n$, $n=1$, 2, 3, \dots:
\begin{equation*}
w(\tau) = 6 C_F \sum_{n=1}^\infty \frac{(n+1)\,(2n)!}{n!\,(n+2)!}
\left(-\frac{1}{\tau}\right)^n
= \frac{1}{2} C_F \tau^2 \left[ 1 - \frac{6}{\tau^2}
- \left(1-\frac{2}{\tau}\right) \sqrt{1+\frac{4}{\tau}} \right]\,.
\end{equation*}
At $\tau<1$, it is the sum~(\ref{Ren:w0}) over the IR renormalons
at $u=n+\frac{1}{2}$ ($n=0$, 1, 2, \dots) and at $u=2$:
\begin{equation*}
\begin{split}
w(\tau) &{}= \frac{1}{2} C_F \left[ \tau^2 - 3 \sum_{n=0}^\infty
\frac{(2n-1)\,(2n-1)!!\,(2n-5)!!}{(2n)!} \frac{\tau^{n+\frac{1}{2}}}{(-4)^n}
\right]\\
&{}= \frac{1}{2} C_F \left[ (2-\tau) \sqrt{\tau(4+\tau)} + \tau^2 \right]
\end{split}
\end{equation*}
(both of these series are easily summed using the Newton binomial expansion).
Finally, we obtain (Fig.~\ref{Ren:Mass})
\begin{equation}
w(\tau) = \frac{1}{2} C_F \left[ (2-\tau) \sqrt{\tau(4+\tau)} + \tau^2
- 6 \theta(\tau-1) \right]\,.
\label{RHos:w}
\end{equation}
At $\tau\to0$, $w(\tau)\sim\sqrt{\tau}$,
and at $\tau\to\infty$, $w(\tau)\sim1/\tau$,
according to the positions of the nearest IR and UV renormalons.

The IR renormalon ambiguity of the on-shell mass is~\cite{BB:94},
from the residue of $S(u)$ at the leading IR renormalon $u=1/2$,
\begin{equation}
\Delta m = 2 C_F e^{5/6} \frac{\Lambda_{\MS}}{\beta_0}\,.
\label{RHos:Dm}
\end{equation}
The meson mass is a measurable quantity, and must be unambiguous.
In HQET, it is an expansion in $1/m$.
Its leading term, $m$, is a short-distance quantity -- a parameter of QCD.
The first correction, $\bar{\Lambda}$,
is a long-distance quantity, determined by the meson structure
at the confinement scale.
However, \MS{} regularization scheme contains no strict momentum cutoffs.
As a result, the on-shell mass $m$ also contains a contribution
from large distances, where perturbation theory is ill-defined.
This produces the IR renormalon ambiguity~(\ref{RHos:Dm}),
which is suppressed by 1/m as compared to the leading term.
Likewise, $\bar{\Lambda}$ contains a contribution from small distances,
which leads to the UV renormalon ambiguity~(\ref{RH:DL}).
They compensate each other in the physical quantity -- the meson mass.
In other words, in \MS{} the separation of the short- and long-distance
contributions is ambiguous, though the full result is not.

This cancellation should hold beyond the large-$\beta_0$ limit.
Therefore~\cite{Be:95},
\begin{equation}
S(u) = \frac{2 C_F N_0'}{\left(\frac{1}{2}-u\right)^{1+\frac{\beta_1}{2\beta_0^2}}}
\left[1 + \mathcal{O}\left(\tfrac{1}{2}-u\right)\right]\,,
\label{RHos:branch}
\end{equation}
where the power is exact.
The coefficients in the perturbative series
\begin{equation*}
\frac{m}{\hat{m}} = 1 + \frac{1}{\beta_0} \sum_{L=1}^\infty
c_L \left(\frac{\alpha_s(\mu_0)}{4\pi}\right)^L
\end{equation*}
at $L\gg1$ are, according to~(\ref{beta:cL}),
\begin{equation*}
c_{n+1} = 2^{1+a} 2 C_F N_0' (2\beta_0)^n (1+a)_n
\left[1 + \mathcal{O}(1/n)\right]\,,\quad
a = \frac{\beta_1}{2\beta_0^2}\,.
\end{equation*}
From the Stirling formula, $\Gamma(n+1+a)=n^a n! [1+\mathcal{O}(1/n)]$,
and we arrive at
\begin{equation}
c_{n+1} = 4 C_F N_0\,n!\,(2\beta_0)^n (2\beta_0 n)^{\frac{\beta_1}{2\beta_0^2}}
\left[1 + \mathcal{O}(1/n)\right]\,.
\label{RHos:cL}
\end{equation}
This result is model-independent.

Our calculation of $T(t)$ also yields
$Z_Q^{\text{os}}=\left[1-T'(0)\right]^{-1}$
at the first order in $1/\beta_0$.
It has the form~(\ref{beta:formF}), (\ref{RHos:N}) with
\begin{equation}
N_Z(\varepsilon,u) = - 2 C_F (3-2\varepsilon) (1-u) (1+u-\varepsilon)
\label{RHos:NZ}
\end{equation}
(see~(\ref{RHos:NT})).
If we retain only negative powers of $\varepsilon$,
we obtain $Z_q(\mu)/\widetilde{Z}_Q(\mu)$
(because $\widetilde{Z}_Q^{\text{os}}=1$).
Therefore,
calculating the corresponding anomalous dimension by~(\ref{beta:gamma}),
we obtain
\begin{equation*}
\gamma_q - \widetilde{\gamma}_Q = 2 C_F \frac{\alpha_s}{4\pi}
\frac{(1+\beta)\left(1+\frac{2}{3}\beta\right)}
{B(2+\beta,2+\beta)\Gamma(3+\beta)\Gamma(1-\beta)}
+ \mathcal{O}\left(\frac{1}{\beta_0^2}\right)\,.
\end{equation*}
This difference is gauge-invariant at the $1/\beta_0$ level;
it agrees with~(\ref{RL:gammaq}), (\ref{RH:gammaQ}).
If we retain terms with $\varepsilon^0$,
we get the finite combination $Z_Q^{\text{os}}\widetilde{Z}_Q(\mu)/Z_q(\mu)$
of the form~(\ref{beta:A});
the corresponding renormalization-group invariant~(\ref{beta:Ahat})
has~\cite{NS:95}
\begin{equation*}
S(u) = - 6 C_F \left[ \frac{\Gamma(u)\Gamma(1-2u)}{\Gamma(3-u)} (1-u^2)
- \frac{1}{2u} \right]\,.
\end{equation*}

\section{Chromomagnetic Interaction}
\label{Sec:RMag}

\begin{figure}[ht]
\begin{center}
\begin{picture}(118,30)
\put(59,18){\makebox(0,0){\includegraphics{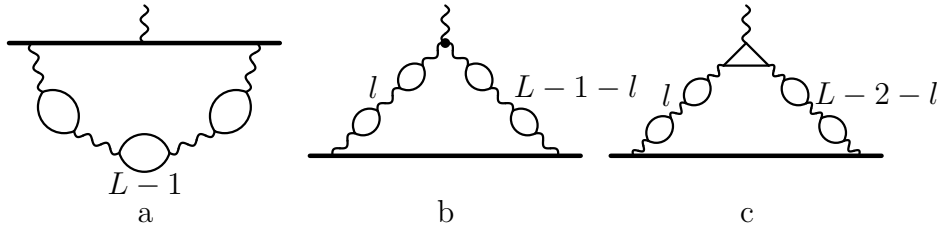}}}
\put(19,5){\makebox(0,0){$L-1$}}
\put(19,0){\makebox(0,0)[b]{a}}
\put(50,18){\makebox(0,0)[r]{$l$}}
\put(68,18){\makebox(0,0)[l]{$L-1-l$}}
\put(59,0){\makebox(0,0)[b]{b}}
\put(89,17){\makebox(0,0)[r]{$l$}}
\put(108,17){\makebox(0,0)[l]{$L-2-l$}}
\put(99,0){\makebox(0,0)[b]{c}}
\end{picture}
\end{center}
\caption{Quark scattering in an external gluon field}
\label{Ren:CMag}
\end{figure}

Now we shall discuss~\cite{GN:97} the chromomagnetic interaction coefficient
$C_m(\mu)$ in the HQET Lagrangian.
It is defined by matching the on-shell scattering amplitudes
in an external chromomagnetic fields in QCD and HQET
at the linear order in the momentum transfer $q$.
All loop diagrams in HQET contain no scale and hence vanish.
The QCD amplitude at the first order in $1/\beta_0$
is given by the $L$-loop diagrams with $L-1$ quark loops
(Fig.~\ref{Ren:CMag}).
The results have the form~(\ref{RHos:N}).
The diagram of Fig.~\ref{Ren:CMag}a is calculated in the standard way,
and gives
\begin{equation*}
N_a(\varepsilon,u) = (2C_F-C_A)
(3+2u-u^2-5\varepsilon+3\varepsilon u-2\varepsilon u^2
+2\varepsilon^2-2\varepsilon^2 u)\,.
\end{equation*}
We should sum the diagrams in~\ref{Ren:CMag}b
in $l$ from 0 to $L-1$.
All terms in the sum are equal,
so that the summation just gives the factor $L=u/\varepsilon$:
\begin{equation*}
N_b(\varepsilon,u) = C_A (5-2u-3\varepsilon)
\frac{u}{\varepsilon}\,.
\end{equation*}
In order to calculate the diagrams in Fig.~\ref{Ren:CMag}c,
we need the triangle quark loop with the linear accuracy in $q$;
it is a combination of one-loop propagator integrals~(\ref{RL:G1}).
Again, all terms in the sum in $l$ from 0 to $L-2$ are equal,
and the summation just gives the factor $L-1$:
\begin{equation*}
N_c(\varepsilon,u) = - C_A
\frac{10-4u-28\varepsilon+9\varepsilon u+23\varepsilon^3-4\varepsilon^2 u
-6\varepsilon^3}{2(1-\varepsilon)}
\left(\frac{u}{\varepsilon}-1\right)\,.
\end{equation*}
Also, we should include the one-loop on-shell quark wave-function
renormalization contribution~(\ref{RHos:NZ}) multiplied by
the Born scattering amplitude (which is just 1).
Finally, we arrive at~\cite{GN:97}
\begin{equation}
\begin{split}
N(\varepsilon,u) &{}= C_F N_F(\varepsilon,u)
+ C_A N_A(\varepsilon,u)\,,\\
N_F(\varepsilon,u) &{}= 4u(1+u-2\varepsilon u)\,,\\
N_A(\varepsilon,u) &{}= \frac{2-u-\varepsilon}{2(1-\varepsilon)}
(2+3u-5\varepsilon-6\varepsilon u+2\varepsilon^2+4\varepsilon^2 u)\,.
\end{split}
\label{RCm:N}
\end{equation}
The sum is regular at the origin $\varepsilon=u=0$,
unlike separate contributions.
It reproduces the leading $\beta_0$ terms
in the two-loop result~\cite{CG:97}.

Now we can easily find the anomalous dimension $\widetilde{\gamma}_m$
and $C_m(\mu)$ with the $1/\beta_0$ accuracy.
The anomalous dimension~(\ref{beta:gamma}) is~\cite{GN:97}
\begin{equation}
\begin{split}
\widetilde{\gamma}_m &{}= C_A \frac{\alpha_s}{2\pi}
\frac{\beta(1+2\beta)\Gamma(5+2\beta)}{24(1+\beta)\Gamma^3(2+\beta)\Gamma(1-\beta)}
+ \mathcal{O}\left(\frac{1}{\beta_0^2}\right)\\
&{}= C_A \frac{\alpha_s}{2\pi} \left[ 1 + \tfrac{13}{6} \beta
- \tfrac{1}{2} \beta^2 + \cdots \right]\,.
\end{split}
\label{RCm:gamma}
\end{equation}
It reproduces the leading-$\beta_0$ term of the two-loop result~\cite{ABN:97,CG:97}
\begin{equation*}
\widetilde{\gamma}_m = C_A \frac{\alpha_s}{2\pi} \left[ 1
+ (13\beta_0-25C_A) \frac{\alpha_s}{24\pi} + \cdots \right]\,.
\end{equation*}
The perturbative series~(\ref{RCm:gamma}) converges at $\beta_0|\alpha_s|<4\pi$.

The renormalization-group invariant $\hat{C}_m$
corresponding to $C_m(\mu)$ (see~(\ref{beta:RGsol}))
has the form~(\ref{beta:Ahat}) with~\cite{GN:97}
\begin{equation}
S(u) = \frac{\Gamma(u)\Gamma(1-2u)}{\Gamma(3-u)}
\left[ 4u(1+u)C_F + \frac{1}{2}(2-u)(2+3u)C_A \right]
- \frac{C_A}{u}\,.
\label{RCm:S}
\end{equation}
The renormalon poles coincide with those in Fig.~\ref{Ren:Heavy}b.
Taking the residue at the leading IR pole $u=1/2$
and comparing with~(\ref{RH:DL}), we obtain
\begin{equation}
\Delta \hat{C}_m = - \left(1+\frac{7}{8}\frac{C_A}{C_F}\right)
\frac{\Delta \bar{\Lambda}}{m}\,.
\label{RCm:DC}
\end{equation}
In physical quantities, such as the mass splitting $m_{B^*}-m_B$,
this IR renormalon ambiguity is compensated by UV renormalon ambiguities
of the matrix elements in the $1/m$ correction.
Detailed investigation of this cancellation allows one
to find the exact nature of the singularity of $S(u)$ at $u=1/2$:
it is a branching point,
a sum of three terms with different fractional powers of $\frac{1}{2}-u$,
where the powers are known exactly,
but the normalizations -- only in the large-$\beta_0$ limit.
The large-$L$ asymptotics of the perturbative series
for $\hat{C}_m$ can be found.
These results have been obtained in~\cite{GN:97}.
We shall not discuss them here,
because they require the use of $1/m^2$ terms in the HQET Lagrangian.
A similar analysis of bilinear heavy--light currents
will be presented in the next Section.

\begin{figure}[ht]
\begin{center}
\begin{picture}(82,33)
\put(41,16.5){\makebox(0,0){\includegraphics{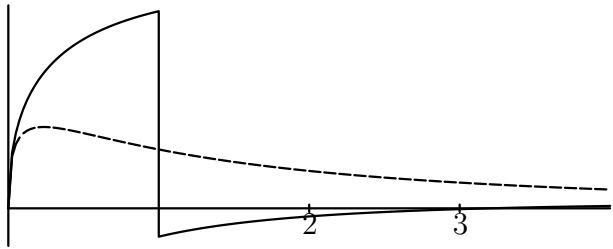}}}
\put(41,1.5){\makebox(0,0)[b]{$2$}}
\put(61,1.5){\makebox(0,0)[b]{$3$}}
\end{picture}
\end{center}
\caption{Distribution functions $w_{\text{F}}$ (dashed line)
and $w_{\text{A}}$ (solid line)}
\label{Ren:wFA}
\end{figure}

We can rewrite $\hat{C}_m$ in the form~(\ref{Ren:Neubert})
with~\cite{GN:97}
\begin{equation}
\begin{split}
w(\tau) &{}= C_F w_F(\tau) + C_A w_A(\tau)\,,\\
w_F(\tau) &{}= 2\tau
\left[\frac{2+4\tau+\tau^2}{\sqrt{\tau(4+\tau)}}-2-\tau\right]\,,\\
w_A(\tau) &{}= \frac{\tau}{4}
\left[\frac{14+5\tau}{\sqrt{\tau(4+\tau)}}-5\right]
- \theta(\tau-1)
\end{split}
\label{RCm:w}
\end{equation}
(Fig.~\ref{Ren:wFA};
these formulae can be derived in the same way as~(\ref{RHos:w})).

\section{Heavy--Light Currents}
\label{Sec:RHL}

Hadronic matrix elements of QCD operators, such as quark currents $j$,
are expanded in $1/m$
\begin{equation}
{<}j{>} = C {<}\widetilde{\jmath}{>}
+ \frac{1}{2m} \sum B_i {<}\widetilde{O}_i{>}
+ \mathcal{O}\left(\frac{1}{m^2}\right)\,,
\label{RHL:exp}
\end{equation}
to separate short-distance contributions --
the matching coefficients $C$, $B_i$, \dots{},
and long-distance ones -- HQET matrix elements
${<}\widetilde{\jmath}{>}$, ${<}\widetilde{O}_i{>}$, \dots{}
The QCD matrix element ${<}j{>}$ contains no renormalon ambiguities,
because the operator $j$ has the lowest dimensionality in its channel.
In schemes without strict separation of large and small momenta,
such as $\overline{\mathrm{MS}}$,
this procedure artificially introduces
infrared renormalon ambiguities in matching coefficients
and ultraviolet renormalon ambiguities in HQET matrix elements.
When calculating matching coefficients $C$, \dots,
we integrate over all loop momenta, including small ones.
Therefore, they contain,
in addition to the main short-distance contributions,
also contributions from large distances,
where the perturbation theory is ill-defined.
They produce infrared renormalon singularities,
%factorially growing contributions to coefficients of the perturbative series,
which lead to ambiguities $\sim\left(\Lambda_{\MS}/m\right)^n$
in the matching coefficients $C$, \dots{}
Similarly, HQET matrix elements of higher-dimensional operators
${<}\widetilde{O}_i{>}$, \dots{}
contain, in addition to the main large-distance contributions,
also contributions from short distances,
which produce several UV renormalon singularities at positive $u$.
They lead to ambiguities of the order $\Lambda_{\MS}^n$
times lower-dimensional matrix elements (e.g., ${<}\widetilde{\jmath}{>}$).
These two kinds of renormalon ambiguities
have to cancel in physical full QCD
matrix elements ${<}j{>}$~(\ref{RHL:exp})~\cite{NS:95}
(see also~\cite{LMW:95}).

\begin{figure}[ht]
\begin{center}
\begin{picture}(56,20)
\put(28,11.5){\makebox(0,0){\includegraphics{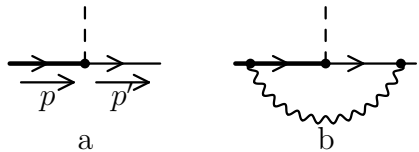}}}
\put(6,5){\makebox(0,0)[b]{$p$}}
\put(16,5){\makebox(0,0)[b]{$p'$}}
\put(11,0){\makebox(0,0)[b]{a}}
\put(43,0){\makebox(0,0)[b]{b}}
\end{picture}
\end{center}
\caption{Proper vertex $\Gamma(p,p')$ of a heavy--light QCD current}
\label{HL:QCD}
\end{figure}

Let's consider the leading QCD/HQET matching coefficient $C_\Gamma(\mu)$
for the heavy--light current with a Dirac matrix $\Gamma$
having the properties
\begin{equation*}
\rlap/v \Gamma = \sigma \Gamma \rlap/v\,,\quad
\sigma = \pm 1\,,
\end{equation*}
and~(\ref{RL:hdef}).
The QCD vertex function $\Gamma(mv,0)$ at one loop (Fig.~\ref{HL:QCD}),
in the Landau gauge, with the gluon denominator raised to a power $n$,
can be calculated, once and for all Dirac matrices,
via $h(d)$~(\ref{RL:hres})~\cite{BG:95}:
\begin{equation*}
a_1(n) = \frac{iC_F}{2(d-1)}
\int \frac{d^d k}{\pi^{d/2}}
\frac{2(d-1)+(d D_2/m^2+4)h-2(D_2/m^2+4)h^2}{D_1 D_2^n}\,.
\end{equation*}
At the first order in $1/\beta_0$,
$\Gamma(mv,0)$ has the form~(\ref{beta:formF}), (\ref{RHos:N})
with~\cite{BG:95}
\begin{equation}
N(\varepsilon,u) = - C_F \left( 2-u-\varepsilon + 2uh - 2h^2 \right)\,.
\label{RHL:N}
\end{equation}
In order to obtain the renormalized matrix element
$\left(Z_Q^{\text{os}}\right)^{1/2}\Gamma(mv,0)$,
we should add $\frac{1}{2}N_Z(\varepsilon,u)$~(\ref{RHos:NZ}).

With the considered accuracy, all loop corrections in HQET vanish.
Retaining negative powers of $\varepsilon$
in $\left(Z_Q^{\text{os}}\right)^{1/2}\Gamma(mv,0)$,
we obtain $Z_j(\mu)/\widetilde{Z}_j(\mu)$.
The corresponding anomalous dimension~(\ref{beta:gamma}) is
\begin{equation}
\gamma_{jn} - \widetilde{\gamma}_j = C_F \frac{\alpha_s}{12\pi}
\frac{2+\beta-2(n-2-\beta)^2+(3+2\beta)(1+\beta)}
{B(2+\beta,2+\beta)\Gamma(3+\beta)\Gamma(1-\beta)}
+ \mathcal{O}\left(\frac{1}{\beta_0^2}\right)\,,
\label{RHL:gamma}
\end{equation}
in agreement with~(\ref{RL:gammaj}) and~(\ref{RHL:gammaj}).
Retaining $\varepsilon^0$ terms, we obtain $C_\Gamma(\mu)$
in the form~(\ref{beta:A}).
The corresponding renormalization-group invariant~(\ref{beta:RGsol})
has the form~(\ref{beta:Ahat}) with~\cite{BG:95}
\begin{align}
S(u) = - C_F \Biggl\{&
\frac{\Gamma(u)\Gamma(1-2u)}{\Gamma(3-u)}
\left[ 5-u-3u^2 + 2u\eta(n-2) - 2(n-2)^2 \right]
\nonumber\\
&{} - \frac{5-2(n-2)^2}{2u} \Biggr\}\,.
\label{RHL:S}
\end{align}
Comparing
the residue at the leading IR renormalon $u=\frac{1}{2}$
with $\Delta \bar{\Lambda}$~(\ref{RH:DL}),
we obtain the ambiguity of the matching coefficient~\cite{BG:95}
\begin{equation}
\Delta C_\Gamma(\mu) = \frac{1}{3}
\left[ \frac{15}{4} + \eta(n-2) - 2(n-2)^2 \right]
\frac{\Delta \bar{\Lambda}}{m}\,.
\label{RHL:DC}
\end{equation}

Matching coefficients for the currents with $\gamma_5^{\text{AC}}$
and $\gamma_5^{\text{HV}}$ have identical $S(u)$ and $\hat{C}_\Gamma$;
they only differ by $K_\gamma(\alpha_s)$ in~(\ref{beta:RGsol}).
For notational convenience, we shall use the $v$ rest frame.
From~(\ref{RHL:N}) and~\cite{BG:95}
\begin{equation*}
Z_P = \frac{C_1}{C_{\gamma_5^{\text{HV}}}}\,,\quad
Z_A = \frac{C_{\gamma^0}}{C_{\gamma_5^{\text{HV}}\gamma^0}}
= \frac{C_{\gamma^i}}{C_{\gamma_5^{\text{HV}}\gamma^i}}
\end{equation*}
we trivially reproduce~(\ref{RL:ZPA}).
Taking into account~\cite{BG:95}
\begin{equation*}
\frac{m}{m(\mu)} = \frac{C_1(\mu)}{C_{\gamma^0}(\mu)}\,,
\end{equation*}
the result~(\ref{RHL:N}) also reproduces the corresponding formula
for $m/m(\mu)$, namely~(\ref{RHos:NT}) with $t=0$.

\begin{figure}[ht]
\begin{center}
\begin{picture}(82,54)
\put(41,28){\makebox(0,0){\includegraphics{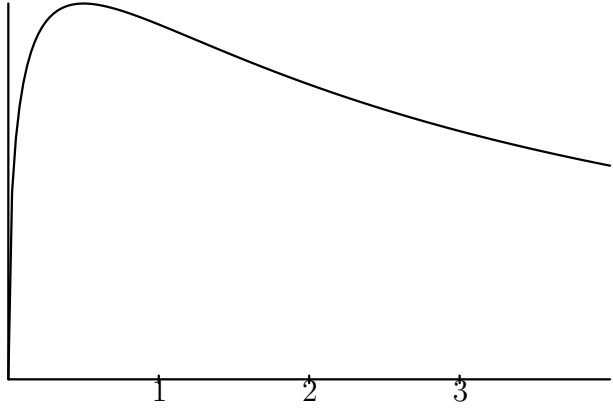}}}
\put(21,0){\makebox(0,0)[b]{1}}
\put(41,0){\makebox(0,0)[b]{2}}
\put(61,0){\makebox(0,0)[b]{3}}
\end{picture}
\end{center}
\caption{Virtuality distribution function for $f_{B^*}/f_B$}
\label{Ren:wf}
\end{figure}

The ratio
\begin{equation*}
\frac{f_{B^*}}{f_B} = \frac{C_{\gamma^i}}{C_{\gamma^0}}
\end{equation*}
is given by~(\ref{beta:Ahat}) with~\cite{NS:95}
\begin{equation}
S(u) = 4 C_F \frac{\Gamma(1+u)\Gamma(1-2u)}{\Gamma(3-u)}\,.
\label{RHL:Sff}
\end{equation}
It can be rewritten in the form~(\ref{Ren:Neubert}).
Summing the series~(\ref{Ren:w0}), (\ref{Ren:winf})
over the residues of $S(u)$, we obtain~\cite{Ne:95}
\begin{equation}
w(\tau) = - \frac{2}{3} C_F \left[ (1+\tau) \sqrt{\tau(4+\tau)}
- \tau(3+\tau) \right]
\label{RHL:w}
\end{equation}
(Fig.~\ref{Ren:wf}).

Now we shall discuss the expansion~(\ref{RHL:exp})
for the matrix elements from $B$ to vacuum
of the currents with $\Gamma=\gamma_5^{\text{AC}}$,
$\gamma_5^{\text{AC}}\gamma^0$ in more detail~\cite{Ne:92,CGM:03}.
The leading HQET current
$\widetilde{\jmath}=\bar{q}\gamma_5^{\text{AC}}\widetilde{Q}$
has the matrix element
\begin{equation}
{<}0|\widetilde{\jmath}(\mu)|B{>} = - i \sqrt{m_B} F(\mu)\,.
\label{Spin0:me0}
\end{equation}
There are 4 dimension-4 HQET operators: 2 local ones and 2 bilocal ones.
The local operators are full derivatives, and their matrix elements
are expressed via $\bar{\Lambda}$ times~(\ref{Spin0:me0}).
The bilocal operators are
\begin{equation}
\widetilde{O}_{jk} = i \int dx\,
\T\left\{\widetilde{\jmath}(0),\widetilde{O}_k(x)\right\}\,,\quad
\widetilde{O}_{jm} = i \int dx\,
\T\left\{\widetilde{\jmath}(0),\widetilde{O}_m(x)\right\}\,,
\label{Spin0:O}
\end{equation}
where $\widetilde{O}_k$, $\widetilde{O}_m$
are the kinetic operator and the chromomagnetic one in the HQET lagrangian.
Their matrix elements are
\begin{equation}
{<}0|\widetilde{O}_{jk}(\mu)|B{>} = - i \sqrt{m_B} F(\mu) G_k(\mu)\,,\quad
{<}0|\widetilde{O}_{jm}(\mu)|B{>} = - i \sqrt{m_B} F(\mu) G_m(\mu)\,.
\label{spin0:me2}
\end{equation}
In the leading logarithmic approximation (LLA),
\begin{equation}
G_k(\mu) = \hat{G}_k
- \bar{\Lambda} \frac{\widetilde{\gamma}^k_0}{2\beta_0}
\log \frac{\alpha_s(\mu)}{4\pi}\,,\quad
G_m(\mu) = \hat{G}_m
\left(\frac{\alpha_s(\mu)}{4\pi}\right)^{\frac{\widetilde{\gamma}_{m0}}{2\beta_0}}
+ \frac{\widetilde{\gamma}^m_0}{\widetilde{\gamma}_{m0}} \bar{\Lambda}\,,
\label{Spin0:LLAG}
\end{equation}
where $\hat{G}_{k,m}$ do not depend on $\mu$,
and $\widetilde{\gamma}^{k,m}$ are mixing anomalous dimensions
of $\widetilde{O}_{jk,jm}$ with local operators, see~\cite{CGM:03}.
The renormalization-group invariant QCD matrix elements
of the pseudoscalar current and the axial one are in LLA
\begin{equation}
\begin{split}
\left\{\begin{array}{c}\hat{f}^P_B\\f_B\end{array}\right\} ={}&
\left(\frac{\alpha_s(m)}{4\pi}\right)^{\frac{\widetilde{\gamma}_{j0}}{2\beta_0}}
\frac{\hat{C}_\Gamma\hat{F}}{\sqrt{m_B}} \Biggl\{ 1 + {}\\
&{} + \frac{1}{2m} \Biggl[
\left(-\frac{\widetilde{\gamma}^k_0}{2\beta_0} \log \frac{\alpha_s(m)}{4\pi}
\pm 1 + \frac{\widetilde{\gamma}^m_0}{\widetilde{\gamma}_{m0}} \right) \bar{\Lambda}
+ \hat{G}_k + \hat{G}_m
\left(\frac{\alpha_s(m)}{4\pi}\right)^{\frac{\widetilde{\gamma}_{m0}}{2\beta_0}}
\Biggr] \Biggr\}\,.
\end{split}
\label{Spin0:fB2}
\end{equation}
The next-to-leading corrections are also known~\cite{BNP:01,CGM:03}.

We are interested in UV renormalon ambiguities
of the matrix elements of $\widetilde{O}_{jk,jm}$.
By dimensional analysis, they are proportional to $\Delta\bar{\Lambda}$
times the matrix element of the lower-dimensional operator $\widetilde{\jmath}$
with the same external states.
We may use quark states instead of hadron ones.
Specifically, we consider transition from an off-shell heavy quark
with residual energy $\omega<0$ to a light quark with zero momentum,
this is enough to ensures the absence of IR divergences.
For $\widetilde{O}_{jk}$, all loop corrections to the vertex function vanish.
The kinetic-energy vertices contain no Dirac matrices,
and we may take $\frac{1}{4}\Tr$ on the light-quark line;
this yields $k^\alpha$ at the vertex,
and the gluon propagator with insertions is transverse.
There is one more contribution:
the matrix element of $\widetilde{\jmath}$ should be multiplied
by the heavy-quark wave-function renormalization $\widetilde{Z}_Q^{1/2}$,
which has an UV renormalon ambiguity~(\ref{RH:DZQ}).
Therefore~\cite{NS:95},
\begin{equation}
\Delta G_{\text{k}}(\mu) = - \frac{3}{2} \Delta \bar{\Lambda}
\label{RHL:DGk}
\end{equation}
(this ambiguity is $\mu$-independent at the first order in $1/\beta_0$).

\begin{figure}[ht]
\begin{center}
\begin{picture}(98,24)
\put(49,13){\makebox(0,0){\includegraphics{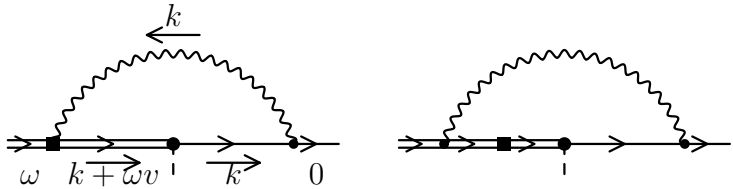}}}
\put(4,0){\makebox(0,0)[b]{$\strut{}\omega$}}
\put(15,0){\makebox(0,0)[b]{$\strut{}k+\omega v$}}
\put(31,0){\makebox(0,0)[b]{$\strut{}k$}}
\put(42,0){\makebox(0,0)[b]{$\strut{}0$}}
\put(23,21){\makebox(0,0)[b]{$\strut{}k$}}
\end{picture}
\end{center}
\caption{Matrix element of $\widetilde{O}_{jk}$}
\label{Ren:O4}
\end{figure}

\begin{figure}[ht]
\begin{center}
\begin{picture}(46,24)
\put(23,13){\makebox(0,0){\includegraphics{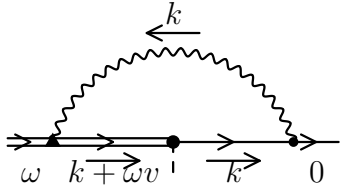}}}
\put(4,0){\makebox(0,0)[b]{$\strut{}\omega$}}
\put(15,0){\makebox(0,0)[b]{$\strut{}k+\omega v$}}
\put(31,0){\makebox(0,0)[b]{$\strut{}k$}}
\put(42,0){\makebox(0,0)[b]{$\strut{}0$}}
\put(23,21){\makebox(0,0)[b]{$\strut{}k$}}
\end{picture}
\end{center}
\caption{Matrix element of $\widetilde{O}_{jm}$}
\label{Ren:O5}
\end{figure}

For $\widetilde{O}_{jm}$, a straightforward calculation
of the diagram in Fig.~\ref{Ren:O5}
gives the bare matrix element
of the usual form~(\ref{beta:formF}), (\ref{RH:F}) with
\begin{equation}
N(\varepsilon,u) = - 6 C_F C_m^0 \frac{\omega}{m}\,.
\label{RHL:N4}
\end{equation}
The renormalization-group invariant matrix element
has the form~(\ref{beta:Ahat}) with $\mu_0=-2\omega e^{-5/6}$ and
\begin{equation}
S(u) = - 6 C_F C_m(\mu_0) \frac{\omega}{m}
\left( \frac{\Gamma(-1+2u)\Gamma(1-u)}{\Gamma(2+u)}
+ \frac{1}{2u} \right)\,.
\label{RHL:S4}
\end{equation}
Taking the residue at the pole $u=1/2$,
we find the UV renormalon ambiguity $C_m(\mu_0)\Delta\bar{\Lambda}/m$
times the matrix element of $\widetilde{\jmath}$,
and we obtain~\cite{NS:95}
\begin{equation}
\Delta G_m(\mu) = 2 \Delta \bar{\Lambda}
\label{RHL:DGm}
\end{equation}
(again, $\mu$-independent at this order).

In the full QCD matrix elements~(\ref{Spin0:fB2}),
the IR renormalon ambiguities~(\ref{RHL:DC})
of the leading matching coefficients $C_\Gamma$
are compensated, at the $1/\beta_0$ order,
by the UV renormalon ambiguities of the subleading matrix elements
$\Delta\bar{\Lambda}$~(\ref{RH:DL})
and $\Delta G_{k,m}$~(\ref{RHL:DGk}), (\ref{RHL:DGm}).
This cancellation must hold beyond the large-$\beta_0$ limit.
The subleading matrix elements are controlled by the renormalization group.
The requirement of cancellation allows one
to investigate the structure of the leading IR renormalon singularity
of $C_\Gamma$~\cite{CGM:03}.

In the large-$\beta_0$ limit,
\begin{equation*}
\Delta \hat{G}_k = - \frac{3}{2} \Delta \bar{\Lambda}\,,\quad
\Delta \hat{G}_m =
\left(2-\frac{\widetilde{\gamma}^m_0}{\widetilde{\gamma}_{m0}}\right)
\Delta \bar{\Lambda}\,,
\end{equation*}
see~(\ref{RHL:DGk}), (\ref{RHL:DGm}), (\ref{Spin0:LLAG}).
In general, they must be equal to $\Lambda_{\MS}$ times some numbers:
\begin{equation}
\Delta \hat{G}_k = - \frac{3}{2} N_1 \Delta_0\,,\quad
\Delta \hat{G}_m =
N_2 \left(2-\frac{\widetilde{\gamma}^m_0}{\widetilde{\gamma}_{m0}}\right) \Delta_0
\label{RHL:Dgen}
\end{equation}
(see~(\ref{RH:DLgen})).
The normalization factors $N_{1,2}$ are only known in the large-$\beta_0$ limit:
\begin{equation*}
N_i = 1 + \mathcal{O}(1/\beta_0)\,;
\end{equation*}
in general, they are just some unknown numbers of order 1.
Using~(\ref{RH:Lambda}), we can represent the UV renormalon ambiguities
of the $1/m$ corrections in~(\ref{Spin0:fB2})
as $\exp[-2\pi/(\beta_0\alpha_s(\mu_0))]$
times a sum of terms with different fractional powers of $\alpha_s(\mu_0)/(4\pi)$.
It is convenient to replace
$\log[\alpha_s(\mu_0)/(4\pi)]\to [(\alpha_s(\mu_0)/(4\pi))^\delta-1]/\delta$,
and take the limit $\delta\to 0$ at the end of calculation.

In order to cancel this ambiguity, we should have the branching point
\begin{equation}
S_\Gamma(u) = \sum_i \frac{r_i}{\left(\frac{1}{2}-u\right)^{1+a_i}}
\label{RHL:Branching}
\end{equation}
instead of a simple pole~(\ref{RHL:S}).
Then (see~(\ref{RH:D}))
\begin{equation}
\Delta \hat{C}_\Gamma = \frac{1}{\beta_0}
\exp\left[-\frac{2\pi}{\beta_0\alpha_s(\mu_0)}\right]
\sum_i \frac{r_i}{\Gamma(1+a_i)}
\left(\frac{\beta_0\alpha_s(\mu_0)}{4\pi}\right)^{-a_i}\,.
\label{RHL:DCgen}
\end{equation}
The requirement of cancellation of the ambiguities in~(\ref{Spin0:fB2})
gives for $\Gamma=1$, $\gamma^0$
\begin{align}
S_\Gamma(u) = {}&
\frac{C_F}{\left(\frac{1}{2}-u\right)^{1+\frac{\beta_1}{2\beta_0^2}}}
\Biggl\{
\left[ - \frac{\widetilde{\gamma}^k_0}{2\beta_0}
\left( \log \frac{\frac{1}{2}-u}{\beta_0}
- \psi\left(1+\frac{\beta_1}{2\beta_0^2}\right) \right)
\pm 1 + \frac{\widetilde{\gamma}^m_0}{\widetilde{\gamma}_{m0}}
\right] N_0'
\nonumber\\
&{} - \frac{3}{2} N_1'
+ \left(2-\frac{\widetilde{\gamma}^m_0}{\widetilde{\gamma}_{m0}}\right) N_2'
\left(\tfrac{1}{2}-u\right)^{\frac{\widetilde{\gamma}_{m0}}{2\beta_0}}
\Biggr\}\,.
\label{RHL:Su1}
\end{align}
The similarly requirement gives for $\Gamma=\gamma^i$, $\gamma^i\gamma^0$
\begin{align}
S_\Gamma(u) = {}&
\frac{C_F}{\left(\frac{1}{2}-u\right)^{1+\frac{\beta_1}{2\beta_0^2}}}
\Biggl\{
\left[ - \frac{\widetilde{\gamma}^k_0}{2\beta_0}
\left( \log \frac{\frac{1}{2}-u}{\beta_0}
- \psi\left(1+\frac{\beta_1}{2\beta_0^2}\right) \right)
+ \frac{1}{3} \left(\pm1-\frac{\widetilde{\gamma}^m_0}{\widetilde{\gamma}_{m0}}\right)
\right] N_0'
\nonumber\\
&{} - \frac{3}{2} N_1'
- \frac{1}{3} \left(2-\frac{\widetilde{\gamma}^m_0}{\widetilde{\gamma}_{m0}}\right) N_2'
\left(\tfrac{1}{2}-u\right)^{\frac{\widetilde{\gamma}_{m0}}{2\beta_0}}
\Biggr\}\,.
\label{RHL:Su2}
\end{align}
Here
\begin{equation*}
N_1' = N_1 \Gamma\left(1+\frac{\beta_1}{2\beta_0^2}\right)
\beta_0^{\frac{\beta_1}{2\beta_0^2}}\,,\quad
N_2' = N_2 \Gamma
\left(1+\frac{\beta_1}{2\beta_0^2}-\frac{\widetilde{\gamma}_{m0}}{2\beta_0}\right)
\beta_0^{\frac{\beta_1}{2\beta_0^2}-\frac{\widetilde{\gamma}_{m0}}{2\beta_0}}
\end{equation*}
(see~(\ref{RH:N0p})).
Corrections $\mathcal{O}\left(\frac{1}{2}-u\right)$
were calculated in~\cite{CGM:03}.
In the large-$\beta_0$ limit, the simple pole behaviour
with~(\ref{RHL:DC}) is reproduced.

The asymptotics of the perturbative coefficients $c^\Gamma_L$ at $L\gg1$
is determined by the renormalon singularity closest to the origin.
Similarly to~(\ref{RHos:cL}), we obtain, for $\Gamma=1$, $\gamma^0$,
\begin{equation}
\begin{split}
c^\Gamma_{n+1} ={}& 2 C_F\,n!\,(2\beta_0)^n\,(2 \beta_0 n)^{\frac{\beta_1}{2\beta_0^2}}
\Biggl[ \left( \frac{\widetilde{\gamma}^k_0}{2\beta_0} \log 2 \beta_0 n
\pm 1 + \frac{\widetilde{\gamma}^m_0}{\widetilde{\gamma}_{m0}} \right) N_0\\
&{} - \frac{3}{2} N_1
+ \left(2-\frac{\widetilde{\gamma}^m_0}{\widetilde{\gamma}_{m0}}\right) N_2
(2 \beta_0 n)^{-\frac{\widetilde{\gamma}_{m0}}{2\beta_0}}\Biggr]\,,
\end{split}
\label{RHL:asy1}
\end{equation}
and for $\Gamma=\gamma^i$, $\gamma^i\gamma^0$,
\begin{equation}
\begin{split}
c^\Gamma_{n+1} ={}& 2 C_F\,n!\,(2\beta_0)^n\,(2 \beta_0 n)^{\frac{\beta_1}{2\beta_0^2}}
\Biggl[ \left( \frac{\widetilde{\gamma}^k_0}{2\beta_0} \log 2 \beta_0 n
+ \frac{1}{3} \left(\pm1
-\frac{\widetilde{\gamma}^m_0}{\widetilde{\gamma}_{m0}}\right) \right) N_0\\
& - \frac{3}{2} N_1
- \frac{1}{3} \left(2-\frac{\widetilde{\gamma}^m_0}{\widetilde{\gamma}_{m0}}\right) N_2
(2 \beta_0 n)^{-\frac{\widetilde{\gamma}_{m0}}{2\beta_0}} \Biggr]\,.
\end{split}
\label{RHL:asy2}
\end{equation}
Corrections $\mathcal{O}(1/n)$ were calculated in~\cite{CGM:03}.

For the ratio $f_{B^*}/f_B$,
the Borel image of the perturbative series is
\begin{equation}
S(u) = \frac{4}{3}\,
\frac{C_F}{\left(\frac{1}{2}-u\right)^{1+\frac{\beta_1}{2\beta_0^2}}}
\left[ \left(1 - \frac{\widetilde{\gamma}^m_0}{\widetilde{\gamma}_{m0}} \right) N_0'
- \left(2-\frac{\widetilde{\gamma}^m_0}{\widetilde{\gamma}_{m0}}\right)
N_2' \left(\tfrac{1}{2}-u\right)^{\frac{\widetilde{\gamma}_{m0}}{2\beta_0}} \right]\,,
\label{RHL:Suff}
\end{equation}
and the asymptotics of the coefficients is
\begin{equation}
c_{n+1} = \frac{8}{3} C_F\,n!\,(2\beta_0)^n
(2 \beta_0 n)^{\frac{\beta_1}{2\beta_0^2}}
\left[ \left( 1 - \frac{\widetilde{\gamma}^m_0}{\widetilde{\gamma}_{m0}} \right) N_0
- \left(2-\frac{\widetilde{\gamma}^m_0}{\widetilde{\gamma}_{m0}}\right) N_2
(2 \beta_0 n)^{-\frac{\widetilde{\gamma}_{m0}}{2\beta_0}} \right]\,.
\label{RHL:asyff}
\end{equation}

\section{Heavy--Heavy Currents}
\label{Sec:RHH}

\begin{figure}[ht]
\begin{center}
\begin{picture}(56,20)
\put(28,11.5){\makebox(0,0){\includegraphics{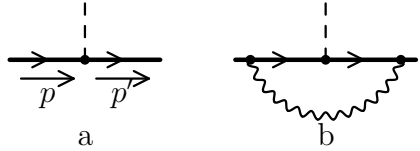}}}
\put(6,5){\makebox(0,0)[b]{$p$}}
\put(16,5){\makebox(0,0)[b]{$p'$}}
\put(11,0){\makebox(0,0)[b]{a}}
\put(43,0){\makebox(0,0)[b]{b}}
\end{picture}
\end{center}
\caption{Proper vertex $\Gamma(p,p')$ of a heavy--heavy QCD current}
\label{HH:QCDvert1}
\end{figure}

First we consider the leading matching coefficients
for the currents $\bar{c}\Gamma b$ at $\vartheta=0$
at $1/\beta_0$ order~\cite{Ne:95a}.
The one-loop vertex $\Gamma(m_b v,m_c v)$ (Fig.~\ref{HH:QCDvert1})
contains integrals
\begin{equation*}
\int \frac{d^d k}%
{\left[m_b^2-(k+m_b v)^2\right]^{n_1}\left[m_c^2-(k+m_c v)^2\right]^{n_2}(-k^2)^n}\,.
\end{equation*}
The denominators are linearly dependent.
We can multiply the integrand by
\begin{equation*}
1 =
\frac{m_b \left[m_c^2-(k+m_c v)^2\right] - m_c \left[m_b^2-(k+m_b v)^2\right]}%
{(m_b-m_c)(-k^2)}\,,
\end{equation*}
thus lowering $n_1$ or $n_2$,
until one of these denominators disappear.
Remaining integrals are single-mass~(\ref{QCDos:M1}).
We have
\begin{equation}
\begin{split}
a_1(n) ={}& C_F
\frac{\Gamma(d-2n-1)\Gamma(-d/2+n+1)}{\Gamma(d-n-1)}
\frac{m_b^{d-2n-1} \Phi(m_c/m_b) - m_c^{d-2n-1} \Phi(m_b/m_c)}{m_b-m_c}\,,\\
\Phi(r) ={}& \frac{d-1}{d-2n-3} r + \frac{2h^2+(d-2n-2)h-d+n+1}{d-n-1}\,.
\end{split}
\label{RHH:a1}
\end{equation}

Adding the on-shell wave-function renormalization~(\ref{RHos:NZ})
for $b$ and $c$, we obtain
\begin{align}
&F(\varepsilon,u) = \left(\frac{\mu^2}{m_b m_c}\right)^u e^{\gamma\varepsilon}
\frac{\Gamma(1+u) \Gamma(1-2u)}{\Gamma(3-u-\varepsilon)}
D(\varepsilon)^{u/\varepsilon-1} N(\varepsilon,u)\,,
\label{RHH:str}\\
&N(\varepsilon,u) = 2 C_F \biggl[ \biggl(
(n-2)^2 - u \eta (n-2) + 2 \varepsilon (n-2) - u \varepsilon \eta
- \frac{4-u^2-4\varepsilon-2u\varepsilon^2}{1+2u} \biggr) R_1
\nonumber\\
&\hphantom{N(\varepsilon,u)=C_F \biggl[\biggr.}
+ (3-2\varepsilon) \frac{1-2u}{1+2u} (1-u-u^2+u\varepsilon) R_0 \biggr]\,,
\label{RHH:Feu}
\end{align}
where
\begin{equation*}
R_0 = \cosh\frac{L}{2}\,,\quad
R_1 = \frac{\sinh\frac{(1-2u)L}{2}}{\sinh\frac{L}{2}}\,,\quad
L = \log\frac{m_b}{m_c}\,,
\end{equation*}
for the on-shell QCD matrix element
(which is equal to the matching coefficient,
because all loop corrections in HQET vanish).
The corresponding anomalous dimension~(\ref{beta:gamma})
reproduces $\gamma_{jn}$~(\ref{RL:gammaj}),
because $\widetilde{\gamma}_J=0$ at $\vartheta=0$.
The function $S(u)$~(\ref{beta:Su}) for the matching coefficient is
\begin{equation}
\begin{split}
S(u) ={}& C_F \biggl\{ 2 \frac{\Gamma(u)\Gamma(1-2u)}{\Gamma(3-u)}
\biggl[ \left( (n-2)^2 - u \eta (n-2) - \frac{4-u^2}{1+2u} \right) R_1\\
&{} + 3 \frac{(1-2u)(1-u-u^2)}{1+2u} R_0 \biggr]
- \frac{(n-1)(n-3)}{u} \biggr\}
\end{split}
\label{RHH:Su}
\end{equation}
with
\begin{equation*}
\mu_0 = e^{-5/6} \sqrt{m_b m_c}
\end{equation*}
(see~(\ref{beta:mu0})).
There is no pole at $u=1/2$,
the leading IR renormalon is at $u=1$.
Therefore, the IR renormalon ambiguity
of the matching coefficients at $\vartheta=0$
is $\sim(\Lambda_{\MS}/m_{b,c})^2/\beta_0$.

For the vector and axial currents~\cite{Ne:95a},
\begin{equation}
\begin{split}
S_{\gamma^0}(u) ={}& 6 C_F \frac{\Gamma(u)\Gamma(1-2u)}{\Gamma(3-u)}
\frac{1-u-u^2}{1+2u}
\Bigl[ - R_1 + (1-2u) R_0 \Bigr]\\
{}={}& 6 C_F \left( \frac{L}{2}\coth\frac{L}{2} - 1 \right)
\left( 1 - \frac{3}{2} u + \cdots \right)\,,\\
S_{\gamma_5^{\text{AC}}\boldsymbol{\gamma}}(u) ={}& 2 C_F
\frac{\Gamma(u)\Gamma(1-2u)}{(1+2u)\Gamma(3-u)}
\Bigl[ - (3-u+u^2) R_1 + (1-2u) (1-u-u^2) R_0 \Bigr]\\
{}={}& C_F \left[ 3 L \coth\frac{L}{2} - 8
- \left( \frac{5}{2} L \coth\frac{L}{2} - 6 \right) u + \cdots \right]\,.
\end{split}
\label{RHH:SVA}
\end{equation}
The matching coefficients don't depend on $\mu$, $\mu'$,
and are given by~(\ref{beta:Ahat}):
\begin{equation}
\begin{split}
H_{\gamma^0} ={}& 6 C_F \frac{\alpha_s(\mu_0)}{4\pi}
\left( \frac{L}{2} \coth\frac{L}{2} - 1 \right)
\left( 1 - \frac{3}{2} \beta + \cdots \right)\,,\\
H_{\gamma_5^{\text{AC}}\boldsymbol{\gamma}} ={}& C_F \frac{\alpha_s(\mu_0)}{4\pi}
\left[ 3 L \coth\frac{L}{2} - 8
- \left( \frac{5}{2} L \coth\frac{L}{2} - 6 \right) \beta + \cdots \right]\,.
\end{split}
\label{RHH:HVA}
\end{equation}

Now we consider the general case $\vartheta\ne0$~\cite{NS:95}.
For a generic Dirac matrix $\Gamma$ satisfying
\begin{equation*}
\gamma_\mu \Gamma \gamma^\mu = 2 h(d) \Gamma\,,
\end{equation*}
there are 4 leading HQET currents in the expansion:
\begin{equation}
\begin{split}
&J = \sum_i H_i \widetilde{J}_i
+ \frac{1}{2 m_b} \sum_i G_i \widetilde{O}_i
+ \frac{1}{2 m_c} \sum_i G'_i \widetilde{O}'_i
+ \mathcal{O}(1/m_{b,c}^2)\,,\\
&J = \bar{c} \Gamma b\,,\quad
\widetilde{J}_i = \overline{\widetilde{c}}_{v'} \Gamma_i \widetilde{b}_v\,,\quad
\Gamma_i = \Gamma\,,\;
\rlap/v\Gamma\,,\;
\Gamma\rlap/v'\,,\;
\rlap/v\Gamma\rlap/v'\,.
\end{split}
\label{HH00:gen}
\end{equation}
The one-loop vertex $\Gamma(m_b v,m_c v')$ in the Landau gauge,
with the gluon denominator raised to a power $n$, is
\begin{equation*}
\Gamma_1 = i C_F g_0^2 \int \frac{d^d k}{(2\pi)^d}
\frac{\gamma_\mu(\rlap/k+m_c\rlap/v'+m_c)\Gamma(\rlap/k+m_b\rlap/v+m_b)\gamma^\mu}%
{(-k^2)^n(-k^2-2m_b\,v\cdot k)(-k^2-2m_c\,v'\cdot k)}\,.
\end{equation*}
Using Feynman parametrization, we have
\begin{gather*}
\Gamma_1 = \frac{i C_F n(n+1)}{\pi^{d/2}}
\int \frac{(1-x-x')^{n-1} dx\,dx'\,d^d k'}{(a^2-k^{\prime2})^{n+2}} \times{}\\
\gamma_\mu(\rlap/k'+m_c(1-x')\rlap/v'-m_b x\rlap/v+m_c)\Gamma
(\rlap/k'+m_b(1-x)\rlap/v-m_c x'\rlap/v'+m_b)\gamma^\mu\,,\\
k' = k + m_b x v + m_c x' v'\,,\\
a^2 = (m_b x v + m_c x' v')^2
= m_b^2 x^2 + m_c^2 x^{\prime2} + 2 m_b m_c x x' \cosh\vartheta\,.
\end{gather*}
We calculate the loop integral
and substitute $x=\xi(1+z)/2$, $x'=\xi(1-z)/2$,
\begin{equation*}
a^2 = m_b m_c \xi^2 a_+ a_-\,,\quad
a_\pm = \cosh\frac{L\pm\vartheta}{2} + z \sinh\frac{L\pm\vartheta}{2}\,.
\end{equation*}
The $\xi$ integration is trivial,
and we obtain $H_i$ having the form~(\ref{beta:formF}),
without the leading 1 for $H_{2,3,4}$;
the functions $F_i(\varepsilon,u)$ have the form~(\ref{RHH:str})
with
\begin{equation}
\begin{split}
N_1(\varepsilon,u) ={}& C_F \Biggl\{ \int_{-1}^{+1} \frac{d z}{(a_+ a_-)^{1+u}}
\biggl[ a_+ a_- h^2 (1-2u) - \frac{(1-z^2)h}{4} u (1-2u)\\
&\qquad{}
+ (1+u-\varepsilon) (2-u-\varepsilon) \cosh\vartheta
+ (\cosh L + z \sinh L) u (2-u-\varepsilon) \biggr]\\
&{} - (3-2\varepsilon) (1-u) (1+u-\varepsilon) (r^u+r^{-u}) \Biggr\}\,,\\
N_2(\varepsilon,u) ={}& - C_F \frac{u}{2} \int_{-1}^{+1} \frac{d z}{(a_+ a_-)^{1+u}}
\left[ \frac{e^L (1+z)^2 h}{2} (1-2u) + (1-z) (2-u-\varepsilon) \right]\,,\\
N_3(\varepsilon,u) ={}& - C_F \frac{u}{2} \int_{-1}^{+1} \frac{d z}{(a_+ a_-)^{1+u}}
\left[ \frac{e^{-L} (1-z)^2 h}{2} (1-2u) + (1+z) (2-u-\varepsilon) \right]\,,\\
N_4(\varepsilon,u) ={}& - C_F \frac{h u (1-2u)}{4} \int_{-1}^{+1}
\frac{d z\,(1-z^2)}{(a_+ a_-)^{1+u}}\,.
\end{split}
\label{RHH:Feu2}
\end{equation}
At $\vartheta=0$, the integrals can be easily calculated,
and $N_1+N_2+N_3+N_4$ reproduces~(\ref{RHH:Feu}).
The anomalous dimension~(\ref{beta:gamma}) corresponding to~(\ref{RHH:Feu2})
is $\gamma_{jn}-\widetilde{\gamma}_J$, see~(\ref{RL:gammaj}), (\ref{RH:gammaJ}).
The functions~(\ref{beta:Su})
\begin{equation*}
S_i(u) = \frac{\Gamma(u)\Gamma(1-2u)}{\Gamma(3-u)} N_i(0,u)
- \frac{N_i(0,0)}{2u}
\end{equation*}
have the leading IR renormalon pole at $u=1/2$,
thus producing the ambiguities~(\ref{Ren:Ambig})
\begin{equation*}
\Delta H_i = - \frac{N_i(0,1/2)}{3C_F} \frac{\Delta\bar{\Lambda}}{\sqrt{m_b m_c}}
\end{equation*}
in the matching coefficients.
It is easy to calculate $N_i(0,1/2)$ using the integrals
\begin{equation*}
\int_{-1}^{+1} \frac{d z}{(a_+ a_-)^{3/2}} =
\frac{4\cosh\frac{L}{2}}{\cosh\vartheta+1}\,,\quad
\int_{-1}^{+1} \frac{z\,d z}{(a_+ a_-)^{3/2}} =
- \frac{4\sinh\frac{L}{2}}{\cosh\vartheta+1}\,.
\end{equation*}
We obtain~\cite{NS:95}
\begin{equation}
\begin{split}
&\Delta H_1 = \left( \frac{1}{\cosh\vartheta+1} - \frac{3}{4} \right)
\left( \frac{1}{m_c} + \frac{1}{m_b} \right) \Delta\bar{\Lambda}\,,\\
&\Delta H_2 = \frac{1}{\cosh\vartheta+1}\,\frac{\Delta\bar{\Lambda}}{2 m_c}\,,\quad
\Delta H_3 = \frac{1}{\cosh\vartheta+1}\,\frac{\Delta\bar{\Lambda}}{2 m_b}\,,\quad
\Delta H_4 = 0\,.
\end{split}
\label{RHH:DH}
\end{equation}
They do not depend on the Dirac matrix $\Gamma$ in the current.
As expected, $\Delta H_1+\Delta H_2+\Delta H_3$ vanishes at $\vartheta=0$.

In matrix elements of QCD currents~(\ref{HH00:gen}),
these IR renormalon ambiguities in the leading matching coefficients $H_i$
must be compensated by UV renormalon ambiguities in matrix elements
of the subleading operators $\widetilde{O}_i$, $\widetilde{O}'_i$.
There are two kinds of subleading operators -- local and bilocal.
First we consider local operators, whose coefficients are completely
fixed by reparametrization invariance~\cite{Ne:93}:
\begin{equation}
\begin{split}
\sum_{\text{local}} G_i \widetilde{O}_i ={}&
\overline{\widetilde{c}}_{v'} \Bigl[
\left( H_1\,\Gamma
+ H_2\,\rlap/v \Gamma
+ H_3\,\Gamma \rlap/v'
+ H_4\,\rlap/v \Gamma \rlap/v'
\right) i \rlap{\hspace{0.2em}/}D
+ 2 i \rlap{\hspace{0.2em}/}D \left(H_2\,\Gamma + H_4\,\Gamma \rlap/v'\right)\\
&{} + 2 \left(  H_1'\,\Gamma
+ H_2'\,\rlap/v \Gamma
+ H_3'\,\Gamma \rlap/v'
+ H_4'\,\rlap/v \Gamma \rlap/v'
\right)  i v'\cdot D
\Bigr] \widetilde{b}_v\,,
\end{split}
\label{HH1m:RI}
\end{equation}
where $H_i'$ are derivatives in the argument $\cosh\vartheta$,
and similarly for $\sum G'_i \widetilde{O}'_i$.
These local operators contain either $D_\mu$ or $\overleftarrow{D}_\mu$.
Let's decompose these derivatives into components in the $(v,v')$ plane
and those orthogonal to this plane.
Projection of $D_\mu$ onto the longitudinal plane is
\begin{equation*}
D_\mu \to
\frac{(v_\mu\,v'\cdot D + v'_\mu\,v\cdot D) \cosh\vartheta
- v_\mu\,v\cdot D - v'_\mu\,v'\cdot D}{\sinh^2\vartheta}\,,
\end{equation*}
and similarly for $\overleftarrow{D}_\mu$.
All operators with longitudinal derivatives can be rewritten,
using equations of motion, as full derivatives of the leading currents $\widetilde{J}_i$.
When we are interested in matrix elements from a ground-state meson
into a ground-state meson, we may replace
\begin{equation*}
i \partial_\mu \widetilde{J}_i \to \bar{\Lambda} (v-v')_\mu \widetilde{J}_i\,.
\end{equation*}
In this case, projecting onto the longitudinal plane means
\begin{equation*}
i D_\mu \to \bar{\Lambda}
\frac{v_\mu \cosh\vartheta - v'_\mu}{\cosh\vartheta+1}\,,\quad
-i \overleftarrow{D}_\mu \to \bar{\Lambda}
\frac{v'_\mu \cosh\vartheta - v_\mu}{\cosh\vartheta+1}\,.
\end{equation*}
The longitudinal part of the local $1/m_{b,c}$ contribution~(\ref{HH1m:RI})
to the QCD matrix element ${<}J{>}$ is easily derived by this substitution.
It, clearly, has an UV renormalon ambiguity proportional to $\Delta\bar{\Lambda}$.
Matrix elements of operators with transverse derivatives
cannot be written as matrix elements of the leading currents $\widetilde{J}_i$
times some scalar factors, they require new independent form factors.
Therefore, they contain no UV renormalon ambiguities,
which should have the form $\Delta\bar{\Lambda}$
times lower-dimensional matrix elements of $\widetilde{J}_i$.
The above derivation is exact
(not only valid in the large-$\beta_0$ limit).
At the first order in $1/\beta_0$, we may replace $H_1\to1$,
$H_{2,3,4}\to0$, $H'_i\to0$.
The contribution of the local subleading operators
to the ambiguity of ${<}J{>}$ in this approximation is~\cite{NS:95}
\begin{equation}
\left( 1 - \frac{1}{\cosh\vartheta+1} \right)
\left( \frac{1}{m_c} + \frac{1}{m_b} \right)
\frac{\Delta\bar{\Lambda}}{2} {<}\widetilde{J}_1{>}
- \frac{1}{\cosh\vartheta+1}
\left( \frac{\Delta\bar{\Lambda}}{2 m_c} {<}\widetilde{J}_2{>}
+ \frac{\Delta\bar{\Lambda}}{2 m_b} {<}\widetilde{J}_3{>} \right)\,.
\label{RHH:Loc}
\end{equation}

\begin{figure}[ht]
\begin{center}
\begin{picture}(92,22)
\put(46,11){\makebox(0,0){\includegraphics{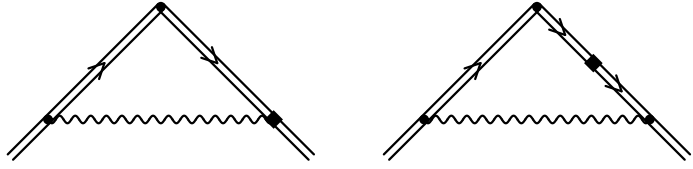}}}
\end{picture}
\end{center}
\caption{Kinetic-energy insertions into the $c$-quark line}
\label{Ren:Kin}
\end{figure}

Now we turn to bilocal subleading operators,
and consider the operator
\begin{equation*}
i \int d x\,\T\left\{\widetilde{J}_1(0),\widetilde{O}_{kc}(x)\right\}
\end{equation*}
with the insertion of the $c$-quark kinetic energy.
It appears in the expansion~(\ref{HH00:gen})
with the coefficient $H_1$.
The one-loop vertex (Fig.~\ref{Ren:Kin}) with the gluon denominator
raised to the power $n$ is
\begin{equation*}
a_1(n) = i \frac{C_F}{2 m_c}
\int \frac{d^d k}{\pi^{d/2}}
\left( k_\bot^\mu - \frac{k_\bot^2 v^{\prime\mu}}{k\cdot v'+\omega'} \right)
\frac{v^\nu}{(k\cdot v+\omega)(k\cdot v'+\omega')(-k^2)^n}
\left( g_{\mu\nu} + \frac{k_\mu k_\nu}{-k^2} \right)\,,
\end{equation*}
where $k_\bot=k-(k\cdot v')\,v'$.
We are interested in the UV renormalon at $u=1/2$;
therefore, to make subsequent formulae shorter,
we shall calculate $F(u)$~(\ref{beta:Fu})
instead of the full function $F(\varepsilon,u)$,
and omit terms regular at $u=1/2$.
We also set $\omega'=\omega$ for simplicity, and obtain
\begin{equation*}
\begin{split}
F(u) ={}& - i \frac{C_F}{2 m_c} u (-2\omega)^{2u} \Biggl[
\int \frac{d^4 k}{\pi^2}
\frac{1}{(-k\cdot v'-\omega) (-k^2)^{1+u}}\\
&{} + \int \frac{d^4 k}{\pi^2}
\frac{\cosh\vartheta}{(-k\cdot v-\omega) (-k\cdot v'-\omega)^2 (-k^2)^u}
+ \cdots \Biggr]\,,
\end{split}
\end{equation*}
where dots mean integrals without linear UV divergences at $u=0$
(and hence having no UV renormalon singularity at $u=1/2$),
and $-2\omega$ plays the role of $m$ in the definition~(\ref{beta:Fu}).

The first integral is trivial (see~(\ref{HQET1:I1})):
\begin{equation*}
- \frac{i}{\pi^2} (-2\omega)^{-1+2u} \int
\frac{d^4 k}{(-k\cdot v'-\omega) (-k^2)^{1+u}} =
2 \frac{\Gamma(-1+2u)\Gamma(1-u)}{\Gamma(1+u)}\,.
\end{equation*}
For the second one,
we use the HQET Feynman parametrization~\cite{BG:91}:
\begin{equation*}
\begin{split}
I ={}& - \frac{i}{\pi^2} (-2\omega)^{-1+2u} \int
\frac{d^4 k}{(-k\cdot v-\omega) (-k\cdot v'-\omega)^2 (-k^2)^u}\\
{}={}& - 8 \frac{i}{\pi^2} (-2\omega)^{-1+2u} \frac{\Gamma(3+u)}{\Gamma(u)}
\int \frac{y'\,d y\,d y'\,d^4 k}%
{\left[-k^2-2yv\cdot k-2y'v'\cdot k-2\omega(y+y')\right]^{3+u}}\\
{}={}& 8 u (-2\omega)^{-1+2u} \int
\frac{y'\,d y\,d y'}%
{\left[y^2+y^{\prime2}+2yy'\cosh\vartheta-2\omega(y+y')\right]^{1+u}}\,.
\end{split}
\end{equation*}
The substitution $y=(-2\omega)\xi(1-z)/2$, $y'=(-2\omega)\xi(1+z)/2$ gives
\begin{equation*}
I = 2 u \int
\frac{\xi^{1-u}\,d\xi\,(1+z)\,d z}%
{\left[\left(\cosh^2\frac{\vartheta}{2}
-z^2\sinh^2\frac{\vartheta}{2}\right)\xi+1\right]^{1+u}}\,.
\end{equation*}
Then
the substitution
$\left(\cosh^2\frac{\vartheta}{2}-z^2\sinh^2\frac{\vartheta}{2}\right)\xi=\eta$
leads to the factored form
\begin{equation*}
I = 2 u
\int_0^\infty \frac{\eta^{1-u} d\eta}{(\eta+1)^{1+u}}
\int_{-1}^{+1} \frac{d z}%
{\left[\cosh^2\frac{\vartheta}{2}-z^2\sinh^2\frac{\vartheta}{2}\right]^{2-u}}\,,
\end{equation*}
where
\begin{equation*}
\int_0^\infty \frac{\eta^{1-u} d\eta}{(\eta+1)^{1+u}} =
\frac{\Gamma(-1+2u)\Gamma(2-u)}{\Gamma(1+u)}\,.
\end{equation*}
Collecting all contributions, we obtain~(\ref{beta:Su})
\begin{equation}
\begin{split}
S(u) ={}& C_F \frac{-2\omega}{m_c}
\frac{\Gamma(-1+2u)\Gamma(1-u)}{\Gamma(1+u)}\times{}\\
&\Biggl[ 1 + u (1-u) \cosh\vartheta
\int_{-1}^{+1} \frac{d z}%
{\left[\cosh^2\frac{\vartheta}{2}-z^2\sinh^2\frac{\vartheta}{2}\right]^{2-u}}
\Biggr] + \cdots
\end{split}
\label{RHH:Su2}
\end{equation}
where dots mean terms regular at $u=1/2$.

The residue at the pole $u=1/2$ can be obtained using
\begin{equation*}
\int_{-1}^{+1} \frac{d z}%
{\left[\cosh^2\frac{\vartheta}{2}-z^2\sinh^2\frac{\vartheta}{2}\right]^{3/2}} =
\frac{1}{\cosh\vartheta+1}\,.
\end{equation*}
The corresponding UV renormalon ambiguity is given by~(\ref{Ren:Ambig})
with $-2\omega$ instead of $m$.
Adding also the external-line renormalization effect $\Delta\widetilde{Z}_c/2$~(\ref{RH:DZQ}),
we obtain
\begin{equation*}
\left( \frac{1}{2} - \frac{1}{\cosh\vartheta+1} \right)
\frac{\Delta\bar{\Lambda}}{2 m_c}\,.
\end{equation*}
The contribution of the bilocal operator with the $b$-quark kinetic energy
contains $m_b$ instead of $m_c$.
Therefore, contribution of all bilocal operators with kinetic-energy insertions
to the ambiguity of ${<}J{>}$ is~\cite{NS:95}
\begin{equation}
\left( \frac{1}{2} - \frac{1}{\cosh\vartheta+1} \right)
\left( \frac{1}{m_c} + \frac{1}{m_b} \right)
\frac{\Delta\bar{\Lambda}}{2} {<}\widetilde{J}_1{>}\,.
\label{RHH:Biloc}
\end{equation}
Matrix elements of bilocal operators with a $c$ or $b$-quark chromomagnetic insertion
cannot be represented as matrix elements of the leading currents $\widetilde{J}_i$
times scalar factors -- they require new independent form factors.
Therefore, they have no UV renormalon ambiguities, which should be equal to
$\Delta\bar{\Lambda}$ times lower-dimensional matrix elements.

Summing the ambiguity $\sum\Delta H_i{<}\widetilde{J}_i{>}$~(\ref{RHH:DH})
of the QCD matrix element ${<}J{>}$ due to the IR renormalon
in the leading matching coefficients $H_i$ at $u=1/2$,
the ambiguity~(\ref{RHH:Loc}) due to the UV renormalon
in the local subleading operators at $u=1/2$,
and the contribution of the bilocal subleading operators~(\ref{RHH:Biloc}),
we see that they cancel, at the first order in $1/\beta_0$,
for any Dirac structure $\Gamma$ of the current $J$~\cite{NS:95}.

\end{document}